# Multimethod Characterization of the French-Pyrenean Valley of Bagnères-de-Bigorre for Seismic-Hazard Evaluation: Observations and Models

by Annie Souriau[1], Emmanuel Chaljub[2], Cécile Cornou[2], Ludovic Margerin[1], Marie Calvet[1], Julie Maury[1], Marc Wathelet[2], Franck Grimaud[3], Christian Ponsolles[3], Catherine Pequegnat[2], Mickaël Langlais[2], and Philippe Guéguen[2]

1. CNRS and Université Paul Sabatier Observatoire Midi-Pyrénées 14 Avenue Edouard Belin 31400—Toulouse, France Annie.Souriau@dtp.obs._mip.fr
2. ISTerre, CNRS-Université Joseph Fourier Grenoble I, IFSTTAR, IRD, Maison des géosciences BP 53X, 38041 Grenoble Cedex, France
3. Observatoire Midi-Pyrénées 57 Avenue d'Azereix 65007-Tarbes, France

**Abstract** A narrow rectilinear valley in the French Pyrenees, affected in the past by damaging earthquakes, has been chosen as a test site for soil response characterization. The main purpose of this initiative was to compare experimental and numerical approaches. A temporary network of 10 stations has been deployed along and across the valley during two years; parallel various experiments have been conducted, in particular ambient noise recording, and seismic profiles with active sources for structure determination at the 10 sites. Classical observables have been measured for site amplification evaluation, such as spectral ratios of horizontal or vertical motions between site and reference stations using direct S waves and S coda, and spectral ratios between horizontal and vertical (H/V) motions at single stations using noise and S-coda records. Vertical shear-velocity profiles at the stations have first been obtained from a joint inversion of Rayleigh wave dispersion curves and ellipticity. They have subsequently been used to model the H/V spectral ratios of noise data from synthetic seismograms, the H/V ratio of S-coda waves based on equipartition theory, and the 3D seismic response of the basin using the spectral element method. General good agreement is found between simulations and observations. The 3D simulation reveals that topography has a much lower contribution to site effects than sedimentary filling, except at the narrow ridge crests. We find clear evidence of a basin edge effect, with an increase of the amplitude of ground motion at some distance from the edge inside the basin and a decrease immediately at the slope foot.

## Introduction

France is considered to be a country of moderate seismicity, if we exclude the French Antilles, where several M > 6 events occurred during the last decade. At the historical timescale, however, the metropolitan territory has experienced strong destructive events causing fatalities, in particular in the Alps, in the Rhine graben, on the French Riviera near Nice, and in the Pyrenees. If similar events were to occur today, they would induce dramatic situations from both human and economical points of view. For this reason, the French government has devoted special effort to evaluate and mitigate seismic risk in both metropolitan territory and the Antilles, and to develop educational programs concerning seismic risk. A six-year Earthquake Plan has been decided by the Ministry of Ecology for the period 2005–2010. In parallel, three urban centers have been selected as pilot sites by the French Accelerometric Network for developing instrumentation and methods for seismic-hazard evaluation: the city of Nice, built on complex geological structures bordering the Mediterranean (Courboulex et al., 2007; Bertrand et al., 2007); the city of Grenoble, located at the convergence of two deep alpine valleys (LeBrun et al., 2001; Cornou et al., 2003; Guéguen et al., 2007); and the two cities of Lourdes and Bagnères-de-Bigorre in the central Pyrenees (Dubos et al., 2003; Souriau et al., 2007). These cities have either a dense permanent population or important sites for tourism.

We present the results obtained at Bagnères-de-Bigorre in the central Pyrenees. The choice of this site was first motivated by its proximity to the seismic sources, with a significant seismicity located less than 20 km away from the city. On the other hand, the Bagnères Valley is nearly rectilinear, with a much simpler geometry than that of the nearby city of Lourdes, where several experiments had been previously conducted (Dubos et al., 2003; Souriau et al., 2007). These conditions are favorable to test different approaches of seismic-hazard evaluation, based both on field experiments and on numerical simulations, and to evaluate the performances and limitations of the different methods.

For this purpose, a variety of experiments have been conducted. In addition to the permanent accelerometric observatory PYBB, a temporary network of 10 accelerometric stations has been deployed along and across the valley. The network operated during two years, thereby providing a database of about 100 events with magnitudes ranging from 2.5 to 5.0. The data have been processed using classical methods such as H=$H_{ref}$ and V=$V_{ref}$ (spectral ratio between the horizontal or vertical signals recorded at one site and those recorded at a reference station on rock), and H/V (ratio of horizontal-to-vertical signals), which give an estimate of the soil response to seismic excitation. In order to model the observations, the structure beneath the stations has been characterized from active surface wave methods. Based on the obtained 1D-velocity profiles, numerical simulations of the 3D response of the valley to various kinds of seismic excitation have been performed and compared with the observations.

After a short description of the geological context, we present a description of the experimental set-up and the data processing. The various experimental approaches to the evaluation of the site response are examined and discussed. The results of 3D numerical simulations are used to elucidate the role of topography and surface heterogeneity in observed amplifications. Finally, we discuss the lessons that may be drawn from these various results, not only for the seismic risk in the Bagnères Valley, but also, from a general point of view, for the performances and limitations of the different experimental and numerical methods used.

## Geological and Tectonic Context

The Pyrenees are an elongated range of mountains extending about 400 km from the Atlantic to the Mediterranean, resulting from the convergence of the Iberian and Eurasian plates during the last 65 m.y., after a rifting episode that opened a shallow sea between the two plates (Choukroune, 1992). The limit between the two plates, the North Pyrenean fault, is located at the northern foot of the range on the French side. It coincides with a sharp Moho jump, the Iberian crust being up to 20 km thicker than the Eurasian one in the central Pyrenees (Hirn et al., 1980; ECORS Pyrenees Team, 1988). The paleo-rift remains as a narrow sedimentary zone north of the North Pyrenean fault (the North Pyrenean zone). South of the North Pyrenean fault, the Paleozoic axial zone contains the highest summits, which reach a 3400 m height in the central part of the range.

The seismic activity is moderate, with about 600 events located each year. It exhibits a general east–west trend, with a greater occurrence of events in the western and central Pyrenees and the presence of several clusters, one of them being located south of Bagnères (Fig. 1). On average only one or two events with M≥5 occur every 10 years (Souriau and Pauchet, 1998; Dubos et al., 2004; Rigo et al., 2005; Sylvander et al., 2008).

The historical seismicity, well-documented back to the fourteenth century (Lambert et al., 1996), reveals a maximum activity in the central part of the range on the French side (Fig. 1), with about 25 events of intensity larger than VII since the beginning of the seventeenth century (in the intensity scale of Medvedev, Sponheuer, and Karnik). Parts of the cities of Lourdes and Bagnères-de-Bigorre were destroyed in 1660 by an event of intensity IX that killed 30 people 17 km southeast of Lourdes and 13 km southwest of Bagnères. Its magnitude is estimated at 6.0–6.1 (Levret et al., 1994). Two other events caused severe damage to these cities in 1750 (intensity VIII) and 1854 (intensity VII).

In a narrow flat valley in the North Pyrenean zone, Bagnères stretches along the south-southeast–north-northwest direction at an altitude of about 500 meters (Fig. 2a). The city is located along the Adour River, which runs to the north-northwest. The Bagnères Valley is filled with fluvial deposits, but imprints of the Quaternary glaciations are visible on the border of the valley. During the Riss episode, the limit of the glaciers was about 3 km north of Bagnères, whereas the Würm glaciation stopped more uphill, 5 km south of the city (Alimen, 1964). Therefore, glacial deposits may be present beneath the fluvial deposits. Drillings performed by the Bureau de Recherches Géologiques et Minières in the valley detected blocks of glacial origin in the southern part of the valley only, and sand and pebble at many places; unfortunately, these drillings do not sample depths larger than 26 m and never reach the bedrock. Bagnères is built at a location where the valley broadens and intercepts a wide crest stretching to the northeast with a road accessing the valley. This crest is covered with Miocene–Pliocene clay and rises 150 m above the valley (Fig. 2b). It is interpreted as a paleotopography preserved between two basins that were affected by a vigorous quaternary erosion (Babault et al., 2005). West of the Adour Valley the mountain rises steeply to a height of 800–1000 m (Fig. 2a). The complex geological nature of the rocks of these mountains, with numerous east–west oriented faults, reflects the tormented history of the structures along the North Pyrenean fault (Fig. 2b).

Bagnères is a small city of 8000 permanent inhabitants, but it is an important tourist center with thermalism, mountain activities, and hotel business induced by the proximity of the pilgrimage city of Lourdes and the Pic-du-Midi astronomical observatory. It also has some industrial activities (electronics, high precision mechanics, cable factory, railway industry, aeronautics). This economical context and the high historical seismicity were additional reasons to choose this city as a pilot site for seismic risk evaluation, an initiative supported by the French Accelerometric Network.

## The Temporary Network

### Description of the Network and Data

A network of 10 accelerometers has been deployed in Bagnères from November 2006 to November 2008 (Table 1), in addition to the permanent station PYBB of the Réseau Accélérométrique Permanent (RAP; Pequegnat et al., 2008). Each station includes a three-component accelerometer Episensor ES-T from Kinemetrics and a Agecodagis Titan3 data logger with a sampling rate of 125 Hz. The recording is continuous for the temporary stations and triggered for the permanent station PYBB, based on a STA/LTA algorithm. Seven of the temporary stations are aligned across the valley (Fig. 2c), with two stations, BBMU and BBCM, at the edge of the valley. The mean distance between these stations is about 500 m. The three remaining stations are set up along the valley axis. Three stations are installed on rock: BBAR, BBMU, PYBB; the other ones are installed on soils.

Most of the stations are located on the floor inside small size buildings of one to three stories or inside individual houses; the resonance frequency of these buildings may be typically of the order of 10 Hz or more (Goel and Chopra, 1998). Only BBAR and BBCM are located in more complex buildings. The frequencies of interest for soil response range from 0.1 to 20 Hz. Hence, the high-frequency spectrum may possibly be perturbed by the resonance of the buildings. This will be clarified later with the aid of horizontal-to-vertical (H/V) ratio measurements. On the other hand, no high building can be found in the vicinity of our stations, the vibrations of which could generate a signal perturbing our measurements (Guéguen et al., 2000; Castellaro and Mulargia, 2010). The permanent accelerometer observatory PYBB is installed inside a historical seismological observatory on a pillar that is anchored to the rock and decoupled from the nearby ground, as was usual practice in the early twentieth century. The instrument and sampling rate are the same as for the temporary network.

The earthquakes recorded by the temporary network are extracted on the basis of the catalog provided by the Seismic Survey Service of the Pyrenees. The catalog includes events with magnitudes as low as 1.0. The accelerometric records are validated and distributed by the Réseau Accélérométrique Permanent (RAP) in Grenoble. All data are high quality, except the north component at the southernmost station BBAS until May 2008, where technical problems have been a posteriori detected. Figure 3 (black traces) shows an example of accelerometric records for an M 4:3 event located 13 km to the west-southwest of Bagnères. These records reveal important variations in the amplitudes and frequency content of the signals, depending on the location of the station. The stations in the valley (BBCA, BBFI, BBAS) recorded a large amplitude, high-frequency signal, whereas the amplitude is low at the foot of the mountain (BBMU). The signals at the stations on the flank of the valley (BBAR, BBHC) have an important low frequency content with a long coda, which appears still more prominent when velocity signals are considered (Fig. 3, gray traces). Site effects are thus very clear from these records, although it is not possible at this stage to know what are the respective contributions of topography and sedimentary filling.

**Characterization of Soil Structure**

A good knowledge of the structure beneath the stations is essential for the interpretation of experimental results and the design of numerical models. Active seismic measurements, such as multichannel analyses of surface waves (MASW; Park et al., 1999) were performed a few tens of meters away from the temporary sites. The experiment consisted in deploying linear profiles from 34.5 to 92 meters in length (Table 2) equipped with 24 Mark Products 4.5 Hz vertical geophones. The waves were generated with a hammer blow on a metallic plate at various offsets from the last geophones in both directions; this offset ranged from one to several times the geophone interspacing. Rayleigh waves dispersion curves for the fundamental and higher modes obtained for each shot points were then processed to get average phase velocities with their confidence level (one standard deviation). Table 2 summarizes the minimum and maximum measured frequencies and wavelengths at each site. Because very inconsistent phase velocities were obtained for different shot points at BBCM, this site has been discarded.

The retrieval of the vertical profiles of shear-wave velocity is performed in two steps. First, fundamental and higher mode Rayleigh waves are inverted using the conditional neighborhood algorithm (Wathelet, 2008), as implemented in the Geopsy package (see Data and Resources). Model parameterization consisted of two or three uniform layers overlaying a half-space, the bottom depth of each layer being defined by a geometrical progression based on the wavelength range, namely from half the minimum measured wavelength to half the maximum measured wavelength. For each site, 20,000 velocity models have been generated. From this model ensemble, we randomly extracted 1000 models having a misfit of 1. The misfit m is defined according to the concept of acceptable solution (Lomax and Snieder, 1994); it has the value m=1 when the calculated dispersion curve is completely inside the uncertainty bounds, and a greater value outside. This allowed us to obtain an ensemble of statistically acceptable models explaining the data within their uncertainty bounds. Superficial shear-wave velocities are varying from site to site (Fig. 4a). At most sites, however, a velocity contrast is detected within the first 10 to 20 m below the surface. This result is consistent with the electrical and gravimeteric prospecting (Perrouty, 2008), which revealed a thickness of quaternary sediments of about 15– 20 m close to BBAS, 25–30 m close to BBFI, and about 42 m close to BBLL.

Deriving the velocity structure at larger depth, especially down to bedrock depth, would require larger aperture acquisition based on active techniques (MASW) or passive techniques such as the frequency-wavenumber (f-k) method (Capon, 1969) or the spatial autocorrelation (SPAC) method (Aki, 1957). Such deployments were not planed in the framework of the present project. However, joint inversion of dispersion and H/V curves has shown its ability to better constrain shear-wave velocity especially at large depths (Fäh et al., 2001; Scherbaum et al., 2003; Parolai et al., 2005; Arai and Tokimatsu, 2005), with a priori assumptions on the energy partition of Rayleigh and Love waves in the noise wave field, or on the ratio between horizontal and vertical loading forces. In order to avoid such prior assumptions, the ellipticity of Rayleigh waves has been measured directly by applying a time-frequency analysis with continuous wavelet transform to the noise wave field (Fäh et al., 2009; Hobiger, 2011). This technique allows us to identify P-SV wavelets in the signal, and ellipticity is estimated by computing the spectral ratio from these wavelets only. In general, reliable measurements of the ellipticity are obtained for the right flank of the ellipticity peak, which carries the most important information on the velocity structure in the intermediate to large depth range (Fäh et al., 2009). Ellipticites were obtained at frequencies lower than those of dispersion curves (Table 2), allowing us to extend the shear-wave velocity profiles at larger depths. For the joint inversion of dispersion and ellipticity, the global misfit was computed by summing single misfit functions obtained from each dataset (ellipticity and dispersion), datasets having similar weight. Weights are indeed useless when considering misfits based on the acceptable solution concept, for which each curve fits the data within its uncertainty bounds with a misfit of one.

Model parameterization is the same as before, except for a layer with linear velocity increase down to 400 m, which has been added above the half-space. The inversion scheme is also the same as before. Figure 5 illustrates the improvement in S-velocity profile when ellipticity and dispersion curves are jointly inverted. Approximate models of shearwave velocity may be derived from the surface down to about 200 m at most sites. Note, however, that when large frequency gaps exist in the ellipticity and dispersion curves measurements, the retrieved S-velocity profiles may be biased at intermediate depths (Hobiger et al., 2010; Hobiger 2011). This is possibly the case at BBFI, BBLL, BBBV, and BBCA sites.

Figure 4b displays the S-velocity profiles obtained in the vicinity of the different temporary stations. At most sites, it has been possible to invert simultaneously phase velocities (fundamental mode and higher modes at a few sites) and ellipticities. At BBMU and BBAS, no stable measurement of ellipticity could be obtained, and the velocity profile is derived from the fundamental Rayleigh wave dispersion curve. The different velocity profiles reveal that, inside the valley, the bedrock is reached at depths of about 100–150 m. The sedimentary layer is thus very thin compared with that observed in Alpine valleys, for example, up to about 1000 m in the Grenoble basin (LeBrun et al., 2001). Moreover, no large variation of the thickness of sediments is observed along the valley. The most significant differences concern the rather strong impedance contrast in the uppermost 20 m, as already mentioned. For numerical modeling based on spectral element method, the following generic S-velocity 1D-profile inside the valley has been derived:

$V_S(m/s) = 200 + 100 \times z^{1/2}$ for z=0–25 m
$V_S(m/s) = 655 + 1.7 \times z$ for z=25–150m
$V_S(m/s) = 2200\ m/s$ for z > 150 m (bedrock)    (1)

where z is depth in meters. The three rock sites BBAR, BBMU, and PYBB display rather different characteristics. BBMU appears as the stiffer site, but the $V_S$ profile could not be retrieved below the depth of 17 m. At BBAR and PYBB, the bedrock velocity is retrieved with a very large uncertainty, with possible $V_S$ values ranging from 1000 to 2500 m/s, thus compatible with those ($V_S \sim 2200$ m/s) found in the middle of the valley, even though the mean value ($V_S \sim 1200$ m/s) may suggest a weathered or fractured bedrock. On the other hand, MASW profiles at BBAR, BBMU, and near PYBB reveal the presence of a soft, thin layer at the surface that may amplify the seismic motion at high frequency. However, this surficial layer does not affect the signal at station PYBB, as the instrument is set up on a pillar anchored in the bedrock. The profile at BBHC reveals a thick, soft layer as expected from the local geology, the station being installed on a crest of Miocene and Pliocene sediments.

# Experimental Site Effect Determination

### Brief Description of the Methods

Three methods are commonly used to determine site effects at one site. The first one is based on the computation of the spectral ratios $H/H_{ref}$ of the horizontal components for local or regional events, at site i, and at a nearby reference site located on bedrock without topography (Borcherdt, 1970). It assumes that the ground motion induced by the earthquake is the same at the sediment-bedrock interface at site i and beneath the surface at the reference site (after correcting the free-surface effect). The $H/H_{ref}$ ratio is thus the transfer function of the sedimentary layers at site i. This is valid only if source radiation and propagation effects can be neglected, thus the distance between source and stations must be large compared with the distance between sites. The method is generally applied to S waves, which are very sensitive to site conditions (in particular to the presence of soft layers) and which are at the origin of most of the damages. It may also be applied to the coda of earthquakes, which results from the scattering of seismic waves in a large volume around the ballistic path from source to station (Phillips and Aki, 1986). In this case, the use of a reference station eliminates statistically most of the contributions of the structures inside this volume. The main difficulty is generally to find a good reference station. The method may also be applied to the vertical component (V/$V_{ref}$), even though vertical acceleration is less important than horizontal acceleration for seismic hazard.

The second method relies on the ratio between horizontal and vertical components of ambient noise at station i (H/V method, Nakamura, 1989). Horizontal-to-vertical measurement of noise is particularly popular because it does not require earthquake recording; only a few tens of minutes of noise record are necessary, and no reference station is necessary. It is thus very easy to implement. However, it leads to results that are sometimes difficult to interpret, because they depend on the way the different components of noise are amplified by the soil structure (Bard, 1999). Difficulties stem from the complexity of ambient noise (which includes Rayleigh waves, Love waves, and body waves) and from its great variability in space and time (Bonnefoy-Claudet, Cotton, and Bard, 2006). However, if high impedance contrasts are present in soil structure, resonances in the uppermost layers generate peaks in the H/V spectral ratios. For a single homogeneous flat layer of thickness h and S-velocity $V_S$ overlying a flat bedrock, a resonance peak is observed at frequency $f=V_S/4h$. The spectral ratio is more complex if impedance contrasts are weak (Malischewsky and Scherbaum, 2004) or if several layers are present. Generally, the frequency of the main peak gives the fundamental resonance frequency, but its amplitude is controlled by the proportion of Love waves and generally underpredicts site amplification (Picozzi et al., 2005; Bonnefoy-Claudet et al., 2008; Haghshenas et al., 2008; Endrun, 2010). At rock site conditions and in the absence of topography, H/V is close to unity in a large frequency domain (Lachet and Bard, 1994; Bard, 1999).

The third method relies on the measurement of energy partitioning of coda waves on horizontal and vertical components (H/V-coda). This method was first proposed by Lermo and Chávez-García (1993), who showed that the site resonance frequency can be estimated by analyzing the H/V ratio of shear waves and their early coda at a single station. They found good agreement between H/V measurements on S waves and the classical spectral ratio results. In a recent study, Margerin et al. (2009) performed an analysis of H/V for coda waves measured at the Pynion Flats observatory in California. They showed that shallow low-velocity layers have a clear impact on the frequency dependence of the H/V ratio in the coda. Just as in the case of noise wave fields, the H/V peak may be related to the resonance frequency of the structure.

The three methods described previously have been applied to data collected at the temporary and permanent stations. Moreover, systematic H/V measurements on noise have been performed to evaluate soil response variability through the whole city using CityShark™, an instrument with a software package Geopsy especially adapted to record urban noise (see Data and Resources; Chatelain et al., 2000; Wathelet et al., 2008). CityShark is connected to a 3DLenhartz velocimeter of period 5 s whose passband and sensitivity are more adapted to record noise than accelerometers. We first present the results for the methods that do not require a reference station (H/V on noise and H/V on S-wave coda), then those where a reference station is necessary (H/$H_{ref}$ on S wave and H/$H_{ref}$ on S coda).

**H/V on Noise: Results**

This method is first considered because it may help to define the reference station, in addition to the MASW results. The processing of the data at the temporary stations follows the procedure described in Souriau et al. (2007). The mean H/V spectral ratio and its confidence level are computed from 30 noise records of 40 s each. The spectra are smoothed using a moving average window whose width is 20% of the central frequency, and the north (N) and east (E) spectra are combined into a single horizontal spectrum according to $H=(|N|^2 + |E|^2)^{1/2}$, assuming the absence of phase coherence between the two components. We have checked that, despite the geometry of the valley, the north to vertical (N/V) and the east to vertical (E/V) spectral ratios are nearly identical. An experiment using the CityShark seismic station has been conducted at the exact location of the temporary stations. For these data, the software package Geopsy provides a quasi-automatic computation of H/V.

*The Influence of the Instrument.* Figure 6a shows the comparison between the results obtained from the accelerometers and from the CityShark velocimeter at BBCA and BBCM. For the accelerometric data, we selected one day (1 February 2008) where the microseismic noise is rather strong. The CityShark experiment has been conducted during a day with rather low microseismic noise. The results obtained at these two stations are well representative of what is obtained at the different stations: there is generally a good coherency of the results at high frequency, but significant discrepancies may be observed at low frequency. This may be explained by the self-noise of the accelerometer compared with the noise level (Strollo et al., 2008), a problem that is not encountered with the CityShark velocimeter. It is illustrated in Figure 6b, which shows the power spectral density of the ambient noise with a clear peak due to the microseismic noise, superimposed on the accelerometer self-noise. The ambient noise is lower than the instrument self-noise at frequencies lower than about 0.1 Hz. Even around 1 Hz the level of ambient noise is sometimes only slightly above the instrumental noise. Thus, our accelerometers are inappropriate to H/V measurements. In what follows, we will thus consider only the CityShark results.

*Effect of the Buildings.* As most stations are set up inside buildings, it is useful to estimate the perturbations they may induce on the records. Figure 6c shows H/V measurements inside the building and immediately outside for a few typical sites, obtained with the CityShark instrument. BBAR and BBCM are located at one end of the basement of recent concrete elongated buildings of 5–6 levels. A small peak at 15 Hz is observed inside the building at BBAR, and at 11 Hz at BBCM. The instrument of BBMU is on the floor in an old, massive stone building of three levels, which does not generate any perturbation. The same conclusion holds at BBBV, BBLL, which are inside individual houses. The spectral ratio perturbation due to the buildings is thus limited to the high-frequency domain of the spectrum, and it appears significantly lower than the spectral ratio variations due to site effects, except perhaps at BBCM. At PYBB, the location of the instrument on a seismological pillar (which is supposed to be anchored to the bedrock) clearly simplifies the site response, as it removes the resonance of the uppermost layers. This shows the efficiency of this observatory installation. Note that using a pillar at the reference station is equivalent to replace soft soil by a stiff structure. It is thus very different from using a reference station in a borehole, where downgoing waves perturb the reference signal, which must be corrected for this effect (Steidl et al., 1996; Assimaki et al., 2008).

*Results at the Different Stations and Choice of the Reference Station.* Figure 6d shows the H/V spectral ratios obtained from CityShark. For the choice of a reference station, we are primarily concerned by the three stations located on rock, BBAR, BBMU, PYBB. We observe a consistent peak at 4.5 Hz at BBAR, possibly due to a topographic effect, and/or to a thin soft layer surface as observed from MASW experiment. At BBMU and PYBB, the spectral ratios are nearly flat and close to unity in the frequency range 0.3– 10 Hz. However, as noted in Figure 3, the amplitudes at BBMU immediately at the slope foot are small compared with those at other stations, suggesting a local topographic or geological effect, as will be discussed later. We will thus choose PYBB as a reference station, keeping in mind that a small peak is present at 17 Hz. This permanent station has in addition the advantage of being maintained at the standard of an observatory, thereby ensuring high-quality data. Note that some stations in the middle of the valley, located on soft soils, have nearly flat H/V spectral ratios, probably because of the absence of strong impedance contrast between stiff sediments and bedrock, as will be discussed later. This illustrates the difficulty of choosing the reference station on the sole basis of the H/V ratio if geological conditions are ignored (Steidl et al., 1996; Cadet et al., 2010).

*H/V Dense Measurements on Noise: Variability of Site Effects in the Valley.* Measurements of H/V spectral ratios on noise are the simplest and least expensive way to have an appraisal of the structure variability throughout the basin, even though only the fundamental resonance frequency could be obtained with this method. They are also of practical interest for risk mitigation, which was one of the motivations of the pilot site experiment. Figure 7a shows the CityShark™ sample points inside the Bagnères basin. They consist of three transverse profiles and one profile along the valley axis. The middle transverse profile and the profile along the valley include the temporary stations.

The H/V spectral ratios are given in Figure 7b. There is no straightforward feature coming out from these results; in particular, it is not possible to draw a map of fundamental resonance frequency, as is sometimes done (e.g., Parolai et al., 2001; Souriau et al., 2007). Even if a change in the noise composition from one measurement to another may be responsible for some variability (due, for example, to the traffic increase at some hours during the experiment), the observed spectral variations mostly reflect the complexity of the structures. In the northern profile where the valley is broad, we observe low frequency peaks close to the valley axis (sites 04, 05, 06), where the sedimentary filling is the thickest. The profile along the valley reveals an important variability of the resonance peaks, with frequency of about 1 Hz at sites 38–39 and about 7 Hz at sites 43–44. As for the H/H$_{ref}$ spectra, it may be due to the increase of sediment thickness downward in the valley. The H/V signal shows some complexity on the flanks of the valley, with sometimes a spectral drop above 10 Hz (sites 02, 08, 09, 19, 24, 29). This is not systematic (e.g., sites 14 and 32 at the foot of the hill, which have a flat spectrum). Sites on the Miocene– Pliocene crest east of the valley (sites 35, 36, 21, 22) exhibit a broad, low frequency peak. We also note a great variability for nearby stations (e.g., 24 and 25). This complexity shows that the detailed geology and geometry of the valley may play an important role at the frequencies we consider; local topography and structures must be taken into account as accurately as possible in the modeling.

**H/V on Coda: Results**

The S coda corresponds to the signal after the direct S-wave arrival, due to wave scattering by crustal heterogeneities. At the distances we consider, the coda regime (defined from energy equipartitioning) is reached only a few seconds after the S arrival, as illustrated in Figure 8. Figure 8a shows a local earthquake record (M 2.9) at station BBAR. Figure 8b shows the typical slow decay of coda energy with time. Coda duration largely exceeds the ballistic propagation time between source and station, as a result of multiple scattering from small-scale heterogeneities in the crust. Figure 8c illustrates the rapid stabilization of the H/V ratio in the coda, typically only a few seconds after the direct shear-wave arrival. This fact is characteristic of coda waves and has previously been reported by Hennino et al. (2001) and Margerin et al. (2009). The return to the noise level is also very clear from the energy analysis: H/V on noise shows very large fluctuations compared with the coda (Fig. 8c).

H/V-coda has been measured on a selection of events at distances larger than 10 km from Bagnères, and with magnitudes larger than 2.5, to get enough energy in the coda (Fig. 9a). The distribution of events around the city exhibits a gap in azimuth to the north and the east (Fig. 1). Moreover, the strong attenuation in the Pyrenees (Drouet et al., 2005) limits the number of remote events that could be recorded. We obtain a total of 83 events with epicentral distances, depths, and magnitudes ranging from 12 to 212 km, 3 to 18 km, and 2.5 to 5.0, respectively (Fig. 9a).

For data processing, we selected an S-wave coda window starting 10 s after the S arrival, with a total duration of 30 s. As for noise records, H/V is defined as the square root of the ratio between the sum of the kinetic energies on the two horizontal components and the kinetic energy on the vertical component. The H/V ratio appears very reproducible and shows little (if any) dependence on source location. This stabilization of the energy ratio in the coda is interpreted as the effect of mode mixing due to multiple scattering. It results in a phenomenon known as equipartition where all the propagation modes, both surface and body waves, get excited by equal energy (Margerin et al., 2009). Results are given in Figure 9b, with 1σ confidence intervals estimated from the variability of the spectra for the different events.

There is overall agreement with H/V on noise, with peaks at the same frequencies and amplitudes ranging roughly from 0.5 to 7. A steady decrease of the H/V ratio at high frequencies (f > 17 Hz) is observed at some stations, whereas it is not observed for noise measurements (e.g., BBBV, BBGA, BBLL). This illustrates that at high frequencies, the wave contents of noise and coda are significantly different. However, the stability and reproducibility of H/V-coda is advantageous over H/V on noise records, which may exhibit large fluctuations.

**$H/H_{ref}$ and $V/V_{ref}$ on S Wave and on S Coda: Results**

*S Waves.* This method has been applied to the same selection of events (Fig. 9a), with magnitudes larger than 2.5 that ensure a good Sn ratio at low frequency and distances to the network larger than 10 km for the reference station method to be valid. S-wave arrivals have been handpicked. A window of length l starting 0.5 s before S has been selected, l increasing with epicentral distance D from l=5 s (for D < 35 km) to l=15 s (for D > 100 km). The spectral ratios have been computed with the same procedure as for H/V (where V is replaced by H at the reference station PYBB). We have first processed independently the north and east components; it shows that there is no variation of horizontal polarization related to the basin north-northwest–south-southeast elongation (@ see Fig. S1, available in the electronic supplement to this paper).

Figure 10a shows the $H/H_{ref}$ spectral ratios obtained after combining the two horizontal components, as well as the $V/V_{ref}$ ratios. In most cases, $H/H_{ref}$ and $V/V_{ref}$ have peaks at the same frequencies and with similar amplitudes, however, $V=V_{ref}$ is generally larger than $H/H_{ref}$ at high frequency. At a few stations (BBLL, BBAR, BBHC), the $V/V_{ref}$ peak is slightly shifted toward high frequency com- pared with the $H/H_{ref}$ peak, as often observed (e.g., Dubos, 2003). From the $H/H_{ref}$ ratios, two interesting results emerge: (1) High ratios (up to 10) are obtained at BBAR and BBHC on the flanks of the valley, even though one of the stations (BBAR) is set up on rock. Note, however, that the resonance frequencies are not the same, about 5 Hz at BBAR and 2.5 Hz at BBHC (on thick soft sediments). By contrast, the stations at the slope foot have smooth spectral ratios. This suggests a possible topographic effect, with amplification at the crest and flank, and amplitude decrease at the slope foot. (2) The spectra of the stations along the valley are highly variable, with a resonance frequency decreasing from south (BBAS and BBFI) to north (BBLL). This corresponds to the downhill increase of the sediment layer thickness, possibly related to the front of the glaciers.

*Coda Waves.* For H/H$_{ref}$ analysis in the coda, we retained the same events and time windows as those used for H/V coda (Fig. 9a). The coda window thus does not generally overlap the window used for S-wave analysis, except for a few remote events. Just as for the S wave ratio, the N/N$_{ref}$ and E/E$_{ref}$ ratios are nearly identical (@ see Fig. S1c, available in the electronic supplement to this paper), denoting the absence of variation in horizontal polarization inside the network. Figure 10b gives the H/H$_{ref}$ spectra, which combines the two horizontal components. They are remarkably consistent with those obtained on S waves. The same conclusion holds for V/V$_{ref}$ (in gray in Fig. 10b). The H/H$_{ref}$ ratios appear, however, slightly smoother and with smaller error bars for coda waves than for S waves, as the result of stabilization of energy partitioning in the coda.

**Comparison of the Different Methods**

Figure 10c shows the intercomparison of the spectral ratios obtained with the different methods. The most prominent feature is the significant difference between the H/V ratios (on noise or coda waves) on one side and the H/H$_{ref}$ ratios (on S waves or coda waves) on the other side. H/H$_{ref}$ exhibits clear peaks at some stations (e.g., BBGA, BBAR) whose amplitude is lower on H/V, sometimes with a nearly constant H/V ratio over a large frequency range (BBMU, BBGA, BBBV). By contrast, the low frequency ratios (for f < 1–2 Hz) are similar.

These differences may hardly be ascribed to the reference station, which is set up on a pillar anchored to the rock and whose H/V ratio is flat up to about 7 Hz. The H/V amplitudes are controlled by the Rayleigh wave ellipticity and by the presence of Love waves that affect the horizontal components. Low H/V may have different origins. (1) Low impedance contrasts between sediments and bedrock combined with low-noise excitation level (Haghshenas et al., 2008). This is probably the case in Bagnères, where rather stiff sediments are overlaying a soft bedrock, leading to moderate shear-wave velocity contrast. (2) Strong lateral variations of the underground structure, especially at sites close to valley edge. In this case, additional lateral interferences of edge-diffracted waves may be generated, which were not present for simple 1D structures where only vertical interferences occur (Uebayashi, 2003; Hagshenas et al., 2008). Differences observed between H/H$_{ref}$ and H/V ratios at some valley sites (BBLL, BBGA) may possibly be attributed to locally diffracted surface waves as previously observed at other sites (e.g., Cornou and Bard, 2003; Bindi et al., 2009).

We have shown the interest of H/V on ambient noise for testing various properties of the sites, for choosing a reference station, and for determining rapidly site effect spatial variability. The comparison of H/V and H=H$_{ref}$ confirms that H/V is at best able to give the fundamental resonance frequency (e.g., at BBLL, BBHC), but that it may strongly underestimate the amplification (Haghshenas et al., 2008), and in some cases (BBGA) H/V may fail in estimating the overall shape of site response (e.g., Cornou and Bard, 2003; Bindi et al., 2009). If earthquakes may be easily recorded, and if a reference station is available, H/H$_{ref}$ on S coda gives the most reliable and most stable results.

# Numerical Models

The confrontation of experimental results with numerical simulations may help to understand the origin of amplification or amplitude decrease at some sites and to specify the role of the various parameters (topography, soils, wave field composition) in site response. Modeling will concern the local response to seismic excitation at the temporary stations, as well as the global response of the basin.

**H/V Modeling with Synthetic Seismograms: A Test of the 1D S-Velocity Profiles**

As the S-velocity models in the vicinity of the temporary stations are known from joint inversion of dispersion curves and ellipticities, it is interesting to check the coherency between these models and the observed H/V measurements on noise records. Because the thickness of the soft layer is small compared with the width of the basin, global resonance of the whole basin is not expected to occur (Bard and Bouchon, 1985); a 1D-modeling relying on the S-velocity profile beneath the station will be appropriate.

In traditional approaches, one considers the response of a stack of plane layers to shear waves arriving at vertical incidence beneath the structures (e.g., Stephenson et al., 2009). A more realistic modeling may, however, be obtained using the complete seismograms. Following Bonnefoy-Claudet, Cornou, et al. (2006), synthetic noise of 13 minutes duration is generated using the discrete wavenumber code developed by

Hisada (1994, 1995), which computes Green's functions due to point sources for viscoelastic horizontally stratified media, using the reflectivity method. Green's functions are then convolved with different source time functions, and synthetic seismograms obtained for each source are summed. In this study, source time functions are approximated by surface and subsurface forces, distributed randomly in space, direction (vertical or horizontal), amplitude, and time. The time function is a delta-like signal with a flat spectrum between 0.2 and 20 Hz. Sources are located at 0.5 m depth and distributed at distances ranging from 5 to 500 m away from the site in order to ensure the excitation of the entire sedimentary column. The modeling has been restricted to the stations for which the structure is well constrained (Fig. 4b). The S-velocity model used is the mean velocity structure previously derived at each site down to the bedrock (at about 100–150 m depth), P velocity is twice S velocity, and $Q_S$ (respectively, $Q_P$) is varying from 15 (respectively, 20) for surficial layers to 100 (respectively, 150) for the deepest structure.

Figure 11 compares the results of the modeling with the H/V spectra of the CityShark experiment obtained near the temporary stations outside the buildings. The agreement between observations and models is satisfying over a large frequency range from 0.2 to 15 Hz. In particular, H/V peak positions at low and high frequencies are rather well retrieved, the high-frequency peak being controlled by the surficial velocity structure. Moreover, the synthetic data do not predict sharp H/V peaks, in agreement with the observations. They generally slightly overpredict the peak amplitudes, suggesting either that the velocity contrasts in the models are slightly stronger than in the real structures or that the proportion between horizontal and vertical forces excitation is not properly modeled.

**Simulation of H/V for Coda Waves**

Similar to the analysis performed with noise records, we use the velocity models to calculate the theoretical H/V ratio in the coda, neglecting the possible role of confinement due to basin edges. In the framework of equipartition theory, we write a modal decomposition of the coda wave field in terms of eigenfunctions of the layered medium under the station. The modal amplitudes are uncorrelated (complex) random variables with zero mean and equal variance. Clearly, the true medium under the station is only approximately layered and also displays small-scale fluctutations superimposed on the background structure. In loose terms, equipartition theory says that, as far as average energetic quantities are concerned, the true medium can be replaced by a laterally averaged medium, and the coda wave field can be represented as a sum of surface and body waves coming from all possible azimuths. With these assumptions, we perform a summation over all modes and compute the kinetic energy on each component of the seismogram. Further details on the computational procedure can be found in Margerin (2009). Note that there is no adjustable parameter in the theory. This is an important difference with noise simulations, where the amount of vertical and horizontal sources may be adjusted to fit the data.

The outcome of the calculations is displayed in Figure 12. In the 0.2–2 Hz frequency band, the agreement between observations and data is satisfactory at most stations. The only notable exception is the station BBLL, which shows a broad peak around 1 Hz. At frequencies higher than 2 Hz, the theoretical calculations may sometimes differ significantly from the measurements. As an example, at BBCA, the theory predicts a peak around 10 Hz, which is absent in the data. At BBAS, the predicted peak is observed in the data but with a smaller amplitude. This comparison validates the gross features of the velocity structure as sensed by low frequency waves. However, the finest details of the observed H/V ratio in the coda are not reproduced by our calculations. Some moderate site effects revealed by coda waves at some stations (BBAR, BBLL, BBHC) are not easily modeled with the available velocity profiles. It would be interesting to adjust the local velocity structure to improve the fit between data and theory. However, solving such an inverse problem goes far beyond the goal of our study.

**Modeling the 3D Response of the Valley with the Spectral Element Method**

In order to quantify the respective contributions of surface topography and basin structure to the observed amplifications, we performed 3D simulations of the response of the Bagnères basin to various seismic excitations using the spectral element method (SEM) (e.g., Komatitsch and Vilotte, 1998; Komatitsch and Tromp, 1999; Komatitsch et al., 2005; Chaljub et al., 2007). If the 3D structure is known with sufficient resolution, the SEM provides an efficient tool to model the variability of the ground motion in space and time by accurately accounting for the complexity of seismic wave propagation in the presence of free-surface topography and 3D heterogeneities, in particular mode conversion from body to surface waves in sedimentary basins (e.g., Chaljub et al., 2010), or from surface to body waves at a nonplanar free-surface

(Komatitsch and Vilotte, 1998). The associated local effects on strong ground motion are of particular interest for seismic risk evaluation.

The response of the structure is investigated in two steps: (1) considering the topography alone and (2) considering both the topography and the filling of the valley by sediments. Two types of sources are investigated: (1) a vertically incident plane shear wave with polarization either in the north–south or east–west direction and (2) a realistic double-couple point source with characteristics close to an M 3.9 event, which occurred on 3 May 2008 to the south-southeast of Bagnères.

*Structure Parametrization and Numerical Implementation.* Figure 13a shows a 3D view of the computational domain used for plane waves, the size of which is about 5 km in the vertical direction and 8 km × 11 km in the east–west and north–south directions, respectively. With the mean crustal structure beneath the Bagnères basin being unknown, we adapted a model previously obtained at the nearby city of Lourdes (Dubos et al., 2003; Table 3). The model is made of homogeneous layers, except the topmost layer, where we imposed a velocity and density gradient from the surface down to 179 m above sea level (Table 3). The base of the computational domain is the planar surface with elevation z=-4451 m, and the free surface is given by a digital elevation map (DEM) with spatial resolution 50 m. Surface elevation inside the domain of computation varies from about z=430m to the northwest to z=1560m to the southwest. From the DEM, we define the surface imprint of the quaternary sediments as the set of points for which the elevation is less than 600 m, and the local slope does not exceed 10% (see the red line showing the obtained basin edge on Fig. 13b). This ad hoc criterion was adjusted in order to have the basin edge separating the BBMU (rock) and BBCA (soil) stations, but it was not possible to find a simple proxy to also include the eastern station BBCM in the sediments.

The basin structure is defined as a single layer with uniform thickness h=150 m, and homogeneous P and S velocities, consistent with the results of the seismic profiles (Table 3). The fundamental resonance frequency of this layer is 1 Hz, in agreement with the observations. The structure of the Miocene sedimentary crest to the east beneath BBHC is not known well enough to be safely introduced in the model. Thus, the comparison between simulations and observations will disregard the eastern slope of the profile (BBCM and BBHC). Throughout the model, the S wave quality factor is chosen to be independent of frequency and to scale with the shear-wave velocity as $Q_S=V_S/10$ ($V_S$ in m/s). Neglecting the bulk attenuation yields the definition of the P-wave quality factor: $Q_P=(3/4)\times(V_P/V_S)^2\times Q_S$.

The 3D computational domain for plane waves (Fig. 13a) is discretized with a mesh of 284,256 spectral elements with polynomial order N=4, which results in 19,356,750 grid points. The mesh is designed to sample the wavelengths with a minimum of 5 grid points (i.e., we use at least one spectral element per minimum wavelength), and it is coarsened with depth following a simple conforming strategy as explained in previous studies (Komatitsch et al., 2004; Chaljub et al., 2007). The minimum element size in the basin is 75 m, which results in accurate simulations for frequencies up to 8 Hz. For each seismic excitation, we compute 30 s of ground displacement, velocity, and acceleration at the 11 temporary and permanent stations (red triangles in Fig. 13b), as well as at 315 virtual stations distributed on a regular grid of size 500 m (yellow triangles in Fig. 13b). When a realistic source is considered, the computational domain is extended laterally and in depth in order to include the focus. The computational domain is 12 km × 12 km × 17 km in the vertical, east–west, and north–south directions, respectively (Fig. 13b). It is discretized with the same grid spacing as the computational domain used for plane waves, which results in 47,331,770 grid points.

The source time function is given by a low-pass filtered Dirac pulse for plane wave excitation and by a low-pass filtered Heaviside pulse for the realistic double-couple source. In each case, the source does not radiate energy for frequencies above 8 Hz, allowing us to compute maps of peak ground displacement, velocity, and acceleration without having to store the full time series at all the surface grid points.

Intrinsic attenuation is modeled using the rheology of the generalized Zener body, which is equivalent to the superposition of standard linear solids (Moczo et al., 2007) and implemented through the introduction of memory variables (see Chaljub et al., 2007, and references therein). We use three relaxation mechanisms to model a constant Q in the frequency range 0.5–8 Hz.

Absorbing boundary conditions are implemented using a simple radiation equation (Lysmer and Kuhlemeyer, 1969; Komatitsch and Vilotte, 1998). This is certainly not the best performing technique, but it is known to be insensitive to Poisson's ratio, a useful property when the absorbing boundaries intersect the sedimentary layers (as is the case here for the northern boundary of the computational domain).

*Results for a Plane Wave.* The seismic source consists in a vertically incident plane S wave with particle displacement polarized either in the north–south direction (thus, nearly parallel to the valley axis) or in the east–west direction. This exercise is mostly academic, however, it may give an idea of the response of the valley to a large remote event (for example, from the south of Spain or from Catalonia), for which S waves arrive with steep incidence beneath the Bagnères basin. Figure 14 shows the maps of peak ground accelerations (PGA) obtained for the two polarizations, when we consider the topography alone (Fig. 14a,c) and when the basin structure is added (Fig. 14b,d). [@ The equivalent maps for peak ground velocities (PGV) and peak ground displacement (PGD) are available in the electronic supplement to this paper; see Figs. S2, S3.] In all maps, there is a contamination of the values along the computational domain boundaries caused by the poor performance of absorbing conditions for vertically incident plane waves travelling at grazing incidence along the lateral boundaries of the domain. Therefore, we avoid interpreting peak values for locations closer than about 2 km from the edges of the computational domain.

The simulations with topography alone clearly show amplifications related to short wavelength crests. The amplifications occur inside the east–west oriented structures for incoming plane wave with north–south polarization and in the north–south oriented structures for east–west polarization. The resonant structures have typically 200–300 m lateral extension, and wave focusing occurs at the top of the crests, whereas defocusing occurs on the slopes. This focusing–defocusing effect is well observed for PGA and PGV; it appears smoother on PGD, as displacement representation favors long wavelength signal (@ see Fig. S3a,c, available in the electronic supplement to this paper). Figure 14a brings to light the numerous east–west oriented structures related to the Pyrenean tectonics (@ see Fig. S2a, available in the electronic supplement to this paper). The PGA, PGV, and PGD inside the Adour basin have medium values and do not exhibit significant variations. We just note a small edge effect with an amplitude increase at the border of the basin, followed by a small amplitude decrease immediately at the mountain foot. It is best observed on the synthetic signal (Fig. 15a, note the small amplitude decrease at BBMU). This effect is more visible for the east–west polarization (Fig. 14c), thanks to the north–south orientation of the valley.

The introduction of the sedimentary layer completely modifies the ground motion pattern, as it is seen on PGA maps (Fig. 14b,d) and on time-series of ground accelerations (Fig. 15b). Whatever the polarization of the incoming wave, both the PGA and the duration of ground motion are seen to increase within the basin (see stations BBCA, BBGA, BBBV in Fig. 15b), as well as PGV and PGD. The most striking feature in the basin is a clear increase, by a factor of 2–3, of the PGA values close to the edges (but slightly offset from the basin edge toward the basin axis), as shown for station BBCA in Figure 15b. This is a manifestation of the so-called basin edge effect, caused by the interference between the incoming shear wave and surface waves diffracted off the basin edge, and first invoked by Kawase (1996) to explain the occurrence of the damage belt during the 1995 Kobe event. The characteristics of the amplification zone and its distance to the basin edge strongly depend on the basin structure (in particular on the geometry of the edge, the thicknesses and velocities of the sedimentary layers) and on the characteristics (in particular, the spectral content) of the incoming signal (Kawase, 1996; Pitarka et al., 1998; Hallier et al., 2008). Note, for example, that the maximum is not at the same distance of the edge for north–south (Fig. 14b) and east–west (Fig. 14d) polarization, an effect easier to observe in the narrow uppermost basin to the south. In addition to the amplification close to the basin edge, we note a systematic PGA decrease immediately outside the basin, at the mountain foot, as shown for station BBMU in Figure 15b (@ see also Fig. S2b,d, available in the electronic supplement to this paper). Such an amplitude decrease outside the narrow amplified zone has been observed in other similar contexts, for example, at Kobe (Pitarka et al., 1998; Hallier et al., 2008) or in the Gubbio basin (Bindi et al., 2009). For the north–south polarization, we also observe in the uppermost basin north–south-oriented stripes distant one to another by about 250 m (Fig. 14b), some amplitude variations resulting likely from stationary resonant modes in the east–west direction. By contrast, the PGD inside the basin appears nearly uniform, due to the prevalence of low frequencies in displacement signal (@ see Fig. S3b,d, available in the electronic supplement to this paper). The SEM thus allows us to obtain a picture of the valley response with a great accuracy and to understand the origin of most of the observed features.

*Results for a Realistic Source.* Next we simulate the response of the Bagnères basin to a local event close to an M 3.9 event that occurred on 3 May 2008 and was well recorded by all the stations of the temporary array. The distance of the epicenter to PYBB is about 13 km, and the focal depth is 11 km. We used a double-couple point source with the same focal mechanism as the real event (Fig. 13b), and the source time function was defined as a low-pass filtered step function. The displacement field radiated by the simulated source has therefore a flat amplitude spectrum up to the maximum frequency of 8 Hz.

Figure 16a,b shows the maps of PGA obtained from the simulations with topography alone and with topography and basin structure, respectively. Similar maps are obtained for PGV and PGD (@ see Figs. S4a,b and S5a,b, respectively, available in the electronic supplement to this paper). Maximal PGA values are around 0.4 g on rock sites close to the source. These high values are mainly due to the use of a source time function with corner frequency $f_c$ = 8 Hz, instead of $f_c$ = 2 to 5 Hz for Pyrenean events of such magnitude (Drouet et al., 2005). The PGA values should therefore not be directly compared with the observations. In the absence of sediments (Fig. 16a), the dominant pattern is due to the combined effects of epicentral distance and radiation pattern. Topographic effects with PGA increase at the top of the crests and PGA decrease immediately below are observed systematically along the narrow crests, in particular to the east of the investigated area. No amplitude variation is observed at the foot of the slopes bordering the basin. The introduction of sediments (Fig. 16b) yields a general PGA increase inside the basin, which is reinforced by the basin edge effect, as well as a clear PGA decrease immediately at the foot of the slopes. The same observations hold for PGV and PGD. On the PGD map, we observe an asymmetry of the basin edge effect, which appears stronger to the east than to the west. It could be a consequence of the relative S-wave incidence with respect to the dip angle of the structures sampled by the waves: the amplitude decreases if the ray arrives parallel to the slope (Bouchon, 1973; Kawase and Aki, 1990; Pedersen et al., 1994).

In order to better quantify the role of the sediments, the spectral ratios obtained for the north–south and east–west components from the simulations with and without sediments are given along three profiles across the valley and at a few additional stations in the higher part of the basin (Fig. 17, with location of the stations used in Fig. 16b). The influence of the sediments almost disappears outside the basin (e.g., at site 130 to the south). By contrast, a strong amplification up to 8 is observed around 1 Hz in the valley. We note the amplification due to the basin edge effect (site 219, for example) and a slight amplitude decrease at high frequency at the foot of the slopes (e.g., at site 245), consistent with the PGA maps. In the frequency range of the simulations, the two horizontal components show similar results outside the basin, with a ratio close to 1. Inside the basin, the amplification is generally slightly larger on the east–west component than on the north–south component, indicating a small effect of the valley orientation.

Figure 18 compares the simulations with the observations at the temporary stations. It shows the recorded east–west ground accelerations (Fig. 18a) and the synthetics obtained with (Fig. 18b) and without (Fig. 18c) sediments. The records have been low-pass filtered below 5 Hz, where the displacement spectrum radiated by the source is supposed to be flat (this filtering explains the disappearance of the P wave). As absolute amplitudes are not valid, the synthetic records have been scaled to the observations by normalizing the simulated signals to the observed one at the reference station PYBB. It has also to be kept in mind that the structures beneath the eastern flank (BBCM and BBHC) could not be modeled. The comparison of simulated signals with records is, however, of great interest to validate the model and to understand the origin of the PGA values.

In the absence of sediments, the dominant feature in the synthetic seismograms is the topographic effect. This is illustrated by the two lower traces in Figure 18c, BB102 being at the top of the crest and BB101 in the valley (see their locations in Fig. 16a). The amplification concerns not only the S wave, but also its coda, which is responsible for the PGA value (unfortunately, there are no observations at these sites). When sediments are added (Fig. 18b), the signal inside the valley becomes more complex than in the absence of sediments. The coda becomes very energetic and can be responsible for the PGA value. By contrast, the direct S wave is generally not significantly amplified. A clear amplitude decrease related to basin edge effect is predicted and observed at the base of the slope at BBMU (Fig. 18a,b). The signal at BBAR remains difficult to explain, as the large amplitude and complex S wave observed at this station are poorly compatible with its location on rock. This suggests either that the rock is strongly weathered (MASW profiles show rather slow shallowest layers that are not modeled here), or that strong wave scattering is induced by the very complex geological nature of the slope, with many lithological contrasts (Fig. 2b) and caves due to karst erosion. The characteristics of the record at the reference station PYBB are, by contrast, in good agreement with rock site conditions.

## Discussion and Conclusions

The Bagnères experiment has been conducted for two reasons: (1) to test experimental and numerical methods in a context that is relatively simple from the geometrical and structural point of view (which thus permits a relatively light instrumentation), (2) to infer some guidelines to mitigate the seismic hazard in the Bagnères Valley, which experienced strong historical damaging events. The main interest of a pilot site for seismic-hazard evaluation is to have control of all the steps from observation to simulation. This implies the deployment of appropriate experiments, specific analyses of the shallow structure, a good understanding of the nature and origin of the observed signals, and finally the confrontation of the observations with numerical simulations. As shown by previous studies, this approach may be very fruitful as soon as a dense seismic array may be set up, for example, the Euro-seistest site at Volvi in Greece (Chávez-García et al., 2000; Manakou et al., 2010), or the San Jose site in California (Hartzell et al., 2003; Hartzell et al., 2010), or the Gubbio basin in central Italy (Bindi et al., 2009).

Classical semiempirical methods of site characterization ($H/H_{ref}$ and $V/V_{ref}$ on S waves and S coda, $H/V$ on noise and on S coda) have been implemented along and across the valley. Both $H/H_{ref}$ and $H/V$ give a very local picture of the soil response, as noted in previous studies (e.g., Stephenson et al., 2009). Contrary to some other studies (e.g., Satoh et al., 2001), we have not considered P-wave spectra, whose interest for seismic risk is minor, nor $H/V$ on S waves, which may be strongly dependent on focal mechanisms. The systematic analysis of the energy partitioning following the S wave reveals that the equipartition regime is reached a few seconds after the S arrival only (Fig. 8). As noted by Chávez-García et al. (2002) and Cornou et al. (2003) in similar contexts (narrow valleys in New Zealand and in the French Alps), the small size of the basin implies that locally generated surface waves appear quickly in the records, so that the direct S-wave window is very short. Figure 10c, which summarizes the experimental results, shows that $H/H_{ref}$ ratios are very similar for S wave and S coda. They give access to a broad response spectrum at the sites, provided that a good reference station could be found. This justifies the interest of these spectra for seismic-hazard assessment. By contrast, the $H/V$ spectra on noise give at best the fundamental resonance frequency without a reliable access to the amplification. The horizontal-to-vertical ratio for S coda is somewhat similar to $H/V$ on noise (Fig. 9b), but it is more stable, due to energy equipartitioning in S coda, whereas noise content may vary with time.

In the Bagnères Valley, the $H/H_{ref}$ spectra reveal the presence of resonance peaks with large amplifications (up to 8) along the valley axis, with a frequency decrease downhill, which is well explained by the valley structure (Fig. 10a). By contrast, $H/V$ on noise appears very difficult to interpret, because some stations with large H=H$_{ref}$ amplifications give a nearly flat $H/V$ spectrum (e.g., BBGA on Fig. 10c). The horizontal-to-vertical ratio on noise may, however, be useful to detect rapid lateral variations in the soil response (Fig. 7), which may sometimes vary drastically at the scale of a few tens of meters only. In the Bagnères experiment, $H/V$ has also been used to detect a possible influence of the buildings where the temporary stations were set up (Fig. 6c); in particular, it illustrates the advantage of the seismological pillar at the PYBB observatory. The choice of this reference station rather than BBMU is also a posteriori justified by the 3D-modeling that reveals an amplitude decrease at the basin edge where BBMU is set up.

The determination of the structure is key to model the basin response to seismic excitation. Thus, it was important to retrieve the structure as deep and as accurately as possible. The use of ellipticity together with wave dispersion, and when possible the higher modes of surface waves, allowed us to reach depths up to about 200 m with profiles of maximum length 90 m, thereby providing a convenient experimental method in mountainous regions or in urban contexts. The S-velocity structure has been checked by modeling $H/V$ spectral ratios for noise and for S coda with two different, though certainly related, modeling approaches (Figs. 11, 12). The agreement between simulations and observed ratios is generally satisfying for noise in the whole frequency band, which validates the MASW results. The results of the modeling for coda waves are more contrasted and suggest that some finer details of the local structure remain to be understood.

The 3D-simulations of the basin response using the spectral element method are very informative, even though the adopted structure is a somewhat simplified version of the real one. A modeling at higher frequency would be necessary to account for the finest details of the structure and of the source spectrum but would require extensive geophysical measurements (array noise techniques, noise correlation, active refraction, and reflection seismics). Despite these limitations, the success of the model is certainly due to the simplicity of the geometry and structure of the basin, compared with other sites (e.g., the Los Angeles basin, Komatitsch et al., 2004). This justifies a posteriori the choice of this valley as a pilot site. In particular, the simulations allowed us to study the relative contributions of topography and sediments to site effects which, from an experimental viewpoint, would only be achievable by setting up very dense surface and borehole instrumentation. In the Bagnères basin, the sedimentary filling is clearly at the origin of the dominant contribution (Figs. 14, 16). Significant topographic amplifications may, however, be observed in some cases along the narrow crests in the mountains; they depend on the wave polarity with respect to crest direction (Fig. 14), a geometrical effect previously observed at other sites (e.g., Lovati et al., 2010; Pischiutta et al., 2010). By contrast, no significant influence of the valley orientation could be observed on peak ground acceleration inside the basin (except some resonances in the uppermost narrow valley, see Fig. 14b and @ Fig. S2b, available in the electronic supplement to this paper). This is because the wavelengths considered are small with respect to the size of the basin, whereas they are comparable to the width of the topographic crests. The numerical simulations also suggest that the high amplifications on the flanks of the valley (close to 8 at BBHC and BBAR, Fig. 10) are probably not of topographic origin, but an effect of local geological characteristics.

The most prominent feature of the amplification maps, as revealed by the 3D-modeling, is the basin edge effect, which predicts an overamplification of the PGA by a factor of about 2 in the basin at some distance of the mountain foot. This distance is about 100–200 m for the moderate events we considered, but could become larger for large magnitude events that excite lower frequencies. A PGA decrease by a factor of about 2 is also observed immediately at the edge of the basin. These predictions are in general good agreement with the observations. Note that without this modeling, it would have been difficult to guess the origin of the very low amplitudes at BBMU (Figs. 3, 18a). It is also interesting to note that, at some stations, the PGA is not due to the direct S wave, but to some energetic coda waves arriving a few seconds later.

Important lessons had been drawn from previous studies devoted to large earthquakes (such as Northridge in 1994 and Kobe in 1995). In regions of moderate seismicity, a light experiment similar to that conducted in Bagnères is an efficient and nonexpensive way to analyze in a systematic way the influence of different parameters describing the valley response to seismic excitation. Our results are mostly based on observations of local small to moderate events and on simulations in a limited frequency range. It is likely that large, regional earthquakes would reveal different soil responses, in particular at low frequency, some effects that would require a much longer and heavier experiment to be investigated. Our study shows, however, that numerical modeling may offer now a good way to tackle such issues at somewhat low cost.

## Data and Resources

Accelerometric data are available at the Réseau Accélérométrique Permanent in Grenoble, France, on request, or at http://www.rap.obs.ujf-grenoble.fr (last accessed March 2011). The topographic map is from the French Institut Géographique National (Saint-Mandé, France); the geological map is from the Bureau de Recherches Géologiques et Minières (Orléans, France). The spectral element simulations were performed on the high-performance computing acilities of the Service Commun de Calcul Intensif de l'Observatoire de Grenoble for the plane wave cases. Simulations for the realistic source were performed at the Commissariat à l'Energie Atomique CCRT (GENCI project 2011046060). They required 290 cores for calculation during 5.5 hours. CitySharkTM, an instrument with a software package Geopsy (http://www.geopsy.org, last accessed March 2011), was used to measure the impact of urban noise on soil response variability through the whole city of Bagnères (Chatelain et al., 2000; Wathelet et al., 2008). Most of the figures in this study have been drawn with the Generic Mapping Tool software (Wessel and Smith, 1991).

## Acknowledgments

We thank the different organizations that support the French Accelerometric Network and the French Ministry of Ecology, which supported this project. Part of the funding was also provided by the European Regional Development Fund (ERDF/FEDER), POCTEFA program, in the frame of the SISPyr (Système d'Information Sismique des Pyrénées) project. We also thank the persons and institutions who hosted stations, the mayor of Bagnères-de-Bigorre, M. Sylvander for extraction of the Digital Elevation Model and focal solution computations, and the Commissariat à l'Energie Atomique (CEA) for the access to its high-performance computing facilities. We also thank two anonymous reviewers and the associate editor, S. Parolai, for helpful comments.

## References

Aki, K. (1957). Space and time spectra of stationary stochastic waves, with special reference to microtremors, Bull. Earthq. Res. Inst. Tokyo 35, 415–457.
Alimen, H. (1964). Le Quaternaire des Pyrénées de la Bigorre, Mém. Carte Géol. France, Paris, Imp. Nat., 394 pp. (in French).
Arai, H., and K. Tokimatsu (2005). S-wave velocity profiling by joint inversion of microtremor dispersion curve and horizontal-to-vertical (H/V) spectrum, Bull. Seismol. Soc. Am. 95, 1766–1778.

Assimaki, D., W. Li, J. H. Steidl, and K. Tsuda (2008). Site amplification and attenuation via downhole array seismogram inversion: A comparative study of the 2003 Miyagi-Oki aftershock sequence, Bull. Seismol. Soc. Am. 98, 301–330.
Azambre, B., F. Crouzel, E.-J. Debroas, J.-C. Soulé, and Y. Ternet (1989). Carte géologique, Notice explicative de la feuille Bagnères-de-Bigorre à 1/50000, Ed. B. R. G. M., Orléans, France (in French).
Babault, J., J. Van den Driessche, S. Bonnet, S. Castelltort, and A. Crave (2005). Origin of the highly elevated Pyrenean peneplain, Tectonics 24, 2010–2019.
Bard, P.-Y. (1999). Microtremor measurements: A tool for site effect estimation? in The Effects of Surface Geology on Seismic Motion, Kudo Irikura, Okada, and Sasatani (Editors), Balkema, Rotterdam, 1251–1279.
Bard, P. Y., and M. Bouchon (1985). The two-dimensional resonance of sediment-filled valleys, Bull. Seismol. Soc. Am. 75, 519–541.
Bertrand, E., A. Duval, M. Castan, and S. Vidal (2007). 3D Geotechnical Soil Model of Nice, France, Inferred from Seismic Noise Measurements, for Seismic Hazard Assessment, American Geophysical Union, Fall Meeting 2007, abstract #NS11D-0798.
Bindi, D., S. Parolai, F. Cara, G. Di Giulio, G. Ferretti, L. Luzi, G. Monachesi, F. Pacor, and A. Rovelli (2009). Site amplifications observed in the Gubbio Basin, Central Italy: Hints for lateral propagation effect, Bull. Seismol. Soc. Am. 99, 741–760.
Bonnefoy-Claudet, S., F. Cotton, and P. Y. Bard (2006). The nature of noise wavefield and its applications for site effects studies: A literature review, Earth Sci. Rev. 79, 205–227.
Bonnefoy-Claudet, S., C. Cornou, P. Y. Bard, F. Cotton, P. Moczo, J. Kristek, and D. Fäh (2006). H/V ratio: A tool for site effects evaluation. Results from 1D noise simulations, Geophys. J. Int. 167, 827–837.
Bonnefoy-Claudet, S., A. Köhler, C. Cornou, M. Wathelet, and P. Y. Bard (2008). Effects of Love waves on microtremor H/V ratio, Bull. Seismol. Soc. Am. 98, 288–300.
Borcherdt, R. D. (1970). Effects of local geology on ground motion near San Francisco Bay, Bull. Seismol. Soc. Am. 60, 29–81.
Bouchon, M. (1973). Effect of topography on surface motion, Bull. Seismol. Soc. Am. 63, 615–632.
Cadet, H., P.-Y. Bard, and A. Rodriguez-Marek (2010). Defining standard rock site: Proposition based on the KiK-net database, Bull. Seismol. Soc. Am. 100, 172–195.
Capon, J. (1969). High-resolution frequency-wavenumber spectrum analysis, Proc. of the IEEE 57, 1408–1419.
Castellaro, F., and F. Mulargia (2010). How far from a building does the ground-motion free-field start? The case of the three famous towers and a modern building, Bull. Seismol. Soc. Am. 100, 2080–2094.
Chaljub, E., D. Komatitsch, J.-P. Vilotte, Y. Capdeville, B. Valette, and G. Festa (2007). Spectral-element analysis in seismology, in Advances in Wave Propagation in Heterogeneous Media, Advances in Geophysics, R. Wu and V. Maupin (Editors), Elsevier Academic Press, Amsterdam, 365–419.
Chaljub, E., P. Moczo, S. Tsuno, P.-Y. Bard, M. Käser, M. Stupazzini, and M. Kristekova (2010). Quantitative comparison of four numerical predictions of 3D ground motion in the Grenoble Valley, France, Bull. Seismol. Soc. Am. 100, 1427–1455.
Chatelain, J. L., P. Guéguen, B. Guillier, J. Fréchet, F. Bondoux, J. Sarrault, P. Sulpice, and J. M. Neuville (2000). CityShark: A user-friendly instrument dedicated to ambient noise (microtremor) recording for site and building response studies, Seismol. Res. Lett. 71, 698–703.
Chávez-García, F. J., J. C. Castillo, and W. R. Stephenson (2002). 3D site effects: A thorough analysis of a high-quality dataset, Bull. Seismol. Soc. Am. 92, 1941–1951.
Chávez-García, F. J., D. Raptakis, K. Makra, and K. Pitilakis (2000). Site effects at Eurosesstest-II. Results from 2D numerical modeling and comparison with observations, Soil Dynam. Earthquake Engin. 19, 23–39.
Choukroune, P. (1992). Tectonic evolution of the Pyrenees, Annu. Rev. Earth Planet. Lett. 20, 143–158.
Cornou, C., and P.-Y. Bard (2003). Site-to-bedrock over 1D transfer function ratio: An indicator of the proportion of edge-generated surface waves? Geophys. Res. Lett. 30, 1453–1457.
Cornou, C., P.-Y. Bard, and M. Dietrich (2003). Contribution of dense array analysis to basin-edge-induced waves identification and quantification: Application to Grenoble basin, French Alps (II), Bull. Seismol. Soc. Am. 93, 2624–2648.
Courboulex, F., C. Larroque, A. Deschamps, C. Kohrs-Sansorny, C. Gélis, J. L. Got, J. Charreau, J. F. Stéphan, N. Béthoux, J. Virieux, D. Brunel, C. Maron, A. M. Duval, J-L. Perez, and P. Mondielli (2007). Seismic hazard on the French Riviera: Observations, interpretations and simulations, Geophys. J. Int. 170, 387–400.
Drouet, S., A. Souriau, and F. Cotton (2005). Attenuation, seismic moment and site effects for week motion events: Application to the Pyrenees, Bull. Seismol. Soc. Am. 95, 1731–1748.

Dubos, N., A. Souriau, C. Ponsolles, and J. F. Fels (2003). Etude des effets de site dans la ville de Lourdes (Pyrénées, France) par la méthode des rapports spectraux, Bull. Soc. Géol. Fr. 174, 33–44 (in French).

Dubos, N., M. Sylvander, A. Souriau, C. Ponsolles, S. Chevrot, J. F. Fels, and S. Benahmed (2004). Analysis of the May 2002 earthquake sequence in the central Pyrenees: Consequences for the evaluation of the seismic risk at Lourdes, France, Geophys. J. Int. 156, 527–540.

ECORS Pyrenees Team (1988). The ECORS deep reflection seismic survey across the Pyrenees, Nature 331, 508–510.

Endrun, B. (2010). Love wave contribution to the ambient vibration H/V amplitude peak observed with array measurements, J. Seismol., doi 10.1007/s10950-010-9191-x.

Fäh, D., F. Kind, and D. Giardini (2001). A theoretical investigation of average H/V ratios, Geophys. J. Int. 145, 535–549.

Fäh, D., M. Wathelet, M. Kristekova, H. B. Havenith, B. Endrun, G. Stamm, V. Poggi, J. Burjanek, and C. Cornou (2009). Using Ellipticity Information for Site Characterization, Network of Research Infrastructures for European Seismology, Deliverable JRA4-D4, EC project number 026130, 54 pp.

Goel, R. K., and A. K. Chopra (1998). Period formulas for concrete shear wall buildings, J. Struct. Eng. ASCE 124, 426–433.

Guéguen, P., P.-Y. Bard, and C. S. Oliveira (2000). Experimental and numerical analysis of soil motions caused by free vibrations of a building model, Bull. Seismol. Soc. Am. 90, 1464–1479.

Guéguen, P., C. Cornou, S. Garambois, and J. Banton (2007). On the limitation of the H/V spectral ratio using seismic noise as an exploration tool. Application to the Grenoble basin (France), Pure Appl. Geoph. 164, 115–134.

Haghshenas, E., P. Y. Bard, and N. TheodulidisSESAME WP04 Team (2008). Empirical evaluation of microtremor H/V spectral ratio, Bull. Earthq. Eng. 6, 75–108.

Hallier, S., E. Chaljub, M. Bouchon, and H. Sekiguchi (2008). Revisiting the basin-edge effect at Kobe during the 1995 Hyogo-Ken Nanbu earthquake, Pure Appl. Geophys. 165, 1751–1760.

Hartzell, S., D. Carver, R. A. Williams, S. Harmsen, and A. Zerva (2003). Site response, shallow shear-wave velocity, and wave propagation at the San Jose, California, dense seismic array, Bull. Seismol. Soc. Am. 93, 443–464.

Hartzell, S., L. Ramirez-Guzman, D. Carver, and Pengcheng Liu (2010). Short baseline variations in site response and wave-propagation effects and their structural causes: Four examples in and around the Santa Clara Valley, California, Bull. Seismol. Soc. Am. 100, 2264–2286.

Hennino, R., N. Trégourès, N. M. Shapiro, L. Margerin, M. Campillo, B. A. van Tiggelen, and R. L. Weaver (2001). Observation of equipartition of seismic waves, Phys. Rev. Lett. 86, 3447–3450.

Hirn, A., M. Daignières, J. Gallart, and M. Vadell (1980). Explosion seismic sounding of throws and dips in the continental Moho, Geophys. Res. Lett. 7, 263–266.

Hisada, Y. (1994). An efficient method for computing Green's functions for a layered half-space with sources and receivers at close depths, Bull. Seismol. Soc. Am. 84, 1456–1472.

Hisada, Y. (1995). An efficient method for computing Green's functions for a layered half-space with sources and receivers at close depths (Part 2), Bull. Seismol. Soc. Am. 85, 1080–1093.

Hobiger, M. (2011). Polarization of surface waves: Characterization, inversion and application to seismic hazard assessment, Ph.D. Thesis, Grenoble University, 301 pp.

Hobiger, M., C. Cornou, N. Le Bihan, B. Endrun, F. Renalier, G. Di Giulio, A. Savvaidis, M. Wathelet, and P.-Y. Bard (2010). Joint inversion of Rayleigh wave ellipticity and spatial autocorrelation measurements, SSA 2010 annual meeting, 21–23 April, Portland, Oregon.

Kawase, H., and K. Aki (1990). Topography effect at the critical SV wave incidence: Possible explanation of damage pattern by the WhittierNarrows, California, earthquake of 1 October 1987, Bull. Seismol. Soc. Am. 80, 1–22.

Kawase, H. (1996). The cause of the damage belt in Kobe: "The basin-edge effect", constructive interference of the direct S wave with the basin-induced diffracted/Rayleigh wave, Seismol. Res. Lett. 67, 25–34.

Komatitsch, D., and J. Tromp (1999). Introduction to the spectral-element method for 3-D seismic wave propagation, Geophys. J. Int. 139, 806–822.

Komatitsch, D., and J.-P. Vilotte (1998). The spectral element method: An efficient tool to simulate the seismic response of 2D and 3D geological structures, Bull. Seismol. Soc. Am. 88, 368–392.

Komatitsch, D., Q. Liu, J. Tromp, P. Süss, C. Stidham, and J. H. Shaw (2004). Simulations of ground motion in the Los Angeles basin based upon the spectral-element method, Bull. Seismol. Soc. Am. 94, 187–206.

Komatitsch, D., S. Tsuboi, and J. Tromp (2005). The spectral-element method in seismology, in Seismic Earth: Array Analysis of Broadband Seismograms, American Geophysical Monograph 157, 205–227.

Lachet, C., and P. Y. Bard (1994). Numerical and theoretical investigations on the possibilities and limitations of Nakamura's technique, J. Phys. Earth 42, 337–397.

Lambert, J., A. Levret-Albaret, M. Cushing, and C. Durouchoux (1996). Mille ans de séismes en France, Ouest Editions, Nantes, France, 70 pp. (in French).

LeBrun, B., D. Hatzfeld, and P.-Y. Bard (2001). Site effects study in urban area: Experimental results in Grenoble, Pure Appl. Geophys. 158, 2543–2557.

Lermo, J., and F. J. Chávez-García (1993). Site effect evaluation using spectral ratios with only one station, Bull. Seismol. Soc. Am. 83, 1574–1594.

Levret, A., J. C. Backe, and M. Cushing (1994). Atlas of macroseismic maps for French earthquakes with their principal characteristics, Nat. Hazards 10, 19–46.

Lomax, A., and R. Snieder (1994). Finding sets of acceptable solutions with a genetic algorithm with application to surface wave group dispersion in Europe, Geophys. Res. Lett. 21, 2617–2620.

Lovati, S., M. Bakalovi, M. Massa, G. Ferretti, F. Pacor, and R. Paolucci (2010). Experimental approach for estimating seismic amplification effects at the top of a ridge and their interpretation on ground motion predictions: The case of Narni (central Italy), Geophys. Res. Abstracts 12, EGU2010-4955-1.

Lysmer, J., and R. L. Kuhlemeyer (1969). Finite dynamic model for infinite media, J. Engin. Mechanics Division, ASCE 95, no. EM4, 859–877.

Malischewsky, P. G., and F. Scherbaum (2004). Love's formula and H/V-ratio (ellipticity) of Rayleigh waves, Wave Motion 40, 57–67.

Manakou, M. V., D. G. Raptakis, F. J. Chávez-García, P. I. Apostolidis, and K. D. Pitilakis (2010). 3D soil structure of the Mygdonian basin for site response analysis, Soil Dynam. Earthquake Eng. 30, 1198–1211.

Margerin, L. (2009). Generalized eigenfunctions of layered elastic media and application to diffuse fields, J. Acoust. Soc. Am. 125, 164–174. Margerin, L., M. Campillo, B. A. van Tiggelen, and R. Hennino (2009). Energy partition of seismic coda waves in layered media: Theory and application to Pinyon Flats Observatory, Geophys. J. Int. 177, 571–585.

Mozco, P., J. O. A. Robertsson, and L. Eisner (2007). The finite-difference time-domain method for modeling of seismic wave propagation, in Advances in Wave Propagation in Heterogeneous Earth, R.-S. Wu and V. Maupin (Editors) Elsevier, Amsterdam, Advances in Geophysics 48, 421–516.

Nakamura, Y. (1989). A method for dynamic characteristics estimation of subsurface using microtremor on the ground surface, Quarterly Report of Railway Technical Research Institute (RTRI), Japan 30, no. 1, 25–33.

Park, C. B., R. D. Miller, and J. Xia (1999). Multi-channel analysis of surface waves, Geophysics 64, 800–808.

Parolai, S., P. Bormann, and C. Milkereit (2001). Assessment of the natural frequency of the sedimentary cover in the Cologne area (Germany) using noise measurements, J. Earthquake Eng. 5, 541–564.

Parolai, S., M. Picozzi, S. M. Richwalski, and C. Milkereit (2005). Joint inversion of phase velocity dispersion and H/V ratio curves from seismic noise recordings using a genetic algorithm, considering higher modes, Geophys. Res. Lett. 32, L01303, doi 10.1029/2004 GL021115.

Pedersen, H. A., F. J. Sánchez-Sesma, and M. Campillo (1994). Three-dimensional scattering by two-dimensional topographies, Bull. Seismol. Soc. Am. 84, 1169–1183.

Pequegnat, C., P. Gueguen, D. Hatzfeld, and M. Langlais (2008). The French Accelerometric network (RAP) and national data center (RAP-NDC), Seismol. Res. Lett. 79, 79–89.

Perrouty, S. (2008). Mesures géophysiques du remplissage sédimentaire Quaternaire dans les vallées de Bagnères-de-Bigorre et d'ArgelèsGazost, M2 report LMTG-University Toulouse III, 30 pp. (in French).

Peterson, J. (1993). Observation and modeling of seismic background noise, U.S. Geol. Surv. Tech. Rept. 93-322, 1–95.

Phillips, W. S., and K. Aki (1986). Site amplification of coda waves from local earthquakes in central California, Bull. Seismol. Soc. Am. 76, 627–648.

Picozzi, M., S. Parolai, and S. M. Richwalski (2005). Joint inversion of H/V ratios and dispersion curves from seismic noise: Estimating the S-wave velocity of bedrock, Geophys. Res. Lett. 32, L11308, doi 10.1029/ 2005GL022878.

Pischiutta, M., G. Cultrera, A. Caserta, and L. Luzi (2010). Topographic effects on the hill of Nocera Umbra, central Italy, Geophys. J. Int. 182, 977–987.

Pitarka, A., K. Irikura, T. Iwata, and H. Sekiguchi (1998). Threedimensional simulation of the near-fault ground motion for the Hyogo-ken Nanbu (Kobe), Japan, earthquake, Bull. Seismol. Soc. Am. 88, 428–440.

Rigo, A., A. Souriau, N. Dubos, M. Sylvander, and C. Ponsolles (2005). Analysis of the seismicity in the central part of the Pyrenees (France), and tectonic implications, J. Seismol. 9, 211–222.

Satoh, T., H. Kawase, and S. Matsushima (2001). Differences between site characteristics obtained from microtremors, S waves, P waves, and codas, Bull. Seismol. Soc. Am. 91, 313–334.

Scherbaum, F., F. K. Hinzen, and M. Ohrnberger (2003). Determination of shallow shear wave velocity profiles in the Cologne, Germany area using ambient vibrations, Geophys. J. Int. 152, 597–612.

Souriau, A., and H. Pauchet (1998). A new synthesis of Pyrenean seismicity and its tectonic implications, Tectonophysics 290, 221–244.

Souriau, A., A. Roullé, and C. Ponsolles (2007). Site effects in the city of Lourdes, France, from H/V measurements. Implications for seismic risk evaluation, Bull. Seismol. Soc. Am. 97, 2118–2136.

Steidl, J. H., A. G. Tumarkin, and R. J. Archuleta (1996). What is a reference site? Bull. Seismol. Soc. Am. 86, 1733–1748.

Stephenson, W. J., S. Hartzell, A. D. Frankel, M. Asten, D. L. Carver, and W. Y. Kim (2009). Site characterization for urban seismic hazard in lower Manhattan, New York City, from microtremor array analysis, Geophys. Res. Lett. 36, L03301.

Strollo, A., S. Parolai, K.-H. Jäckel, S. Marzotari, and D. Bindi (2008). Suitability of short-period sensors for retrieveing reliable H/V peaks for frequencies less than 1 Hz, Bull. Seismol. Soc. Am. 98, 671–681.

Sylvander, M., A. Souriau, A. Rigo, A. Tocheport, J.-P. Toutain, C. Ponsolles, and S. Benahmed (2008). The 2006 November, $M_L$ 50 earthquake near Lourdes (France): New evidence for NS extension across the Pyrenees, Geophys. J. Int. 175, 649–664.

Uebayashi, H. (2003). Extrapolation of irregular subsurface structures using the horizontal-to-vertical spectral ratio of long-period microtremors, Bull. Seismol. Soc. Am. 93, 570–582.

Wathelet, M. (2008). An improved neighborhood algorithm: Parameter conditions and dynamic scaling, Geophys. Res. Lett. 35, L09301, doi 10.1029/2008 GL033256.

Wathelet, M., D. Jongmans, M. Ohrnberger, and S. Bonnefoy-Claudet (2008). Array performances for ambient vibrations on a shallow structure and consequences over VS inversion, J. Seismol. 12, 1–19.

Wessel, P., and W. H. F. Smith (1991). Free software helps map and display data, Eos Trans. AGU 72, 441.

Table 1
Stations Used

| Code | Station Location | Latitude (°N)* | Longitude (°E)* | Altitude (m) |
|---|---|---|---|---|
| BBAR | Hospital (Arbizon) | 43.0609 | 0.1431 | 630 |
| BBMU | Salies Museum | 43.0627 | 0.1462 | 550 |
| BBCA | School (Carnot) | 43.0638 | 0.1474 | 550 |
| BBGA | Social center (Gambetta) | 43.0678 | 0.1516 | 545 |
| BBBV | Vacation center (Bonvouloir) | 43.0688 | 0.1536 | 550 |
| BBCM | Senior center (Castelmouly) | 43.0720 | 0.1588 | 555 |
| BBHC | Private house (Haut-de-la-Côte) | 43.0786 | 0.1633 | 625 |
| BBLL | Private house (Lotissement de Laître) | 43.0773 | 0.1426 | 545 |
| BBFI | Centre Laurent Fignon hotel | 43.0558 | 0.1575 | 565 |
| BBAS | Private house (Asté) | 43.0471 | 0.1615 | 600 |
| PYBB | Bagnères Observatory | 43.0586 | 0.1489 | 560 |

*Coordinates in the WGS84 system.

Table 2
MASW Profile Characteristics

|  | Profile Length (m) | Minimum Wavelength (m) | Maximum Wavelength (m) | Minimim Frequency (Hz) | Maximum Frequency (Hz) | Ellipticity Frequency Range (Hz) |
|---|---|---|---|---|---|---|
| BBFI | 69 | 9.6 | 60 | 8.9 | 29.5 | 1.4–1.8 |
| PYBB | 57.5 | 3.6 | 55.3 | 15.1 | 44.9 | 10.7–14.2 |
| BBAR | 69 | 3.4 | 33.7 | 13.1 | 59 | 6.3–10.3 |
| BBLL | 92 | 5.7 | 45.3 | 12.9 | 49.2 | 1.4–1.7 |
| BBGA | 69 | 6.6 | 48.6 | 9.6 | 39.9 | 1.7–6.13 |
| BBBV | 57.5 | 11 | 61.3 | 11.7 | 28.1 | 1.5–2.4 |
| BBCM | 57.5 | – | – | – | – | – |
| BBAS | 92 | 7.3 | 73.8 | 13.1 | 40 | – |
| BBCA | 34.5 | 3.7 | 20 | 23 | 48.3 | 1.6–7 |
| BBMU | 34.5 | 3.8 | 34.7 | 16.1 | 48.2 | – |
| BBHC | 57.5 | 7.9 | 45.9 | 6.2 | 26.2 | 2.8–4.4 |

Table 3
Model for Spectral Element Method Simulations

| $z$ (m) Above Sea Level | $V_S$ (m/s) | $V_P$ (m/s) | Density (kg/m$^3$) |
|---|---|---|---|
| Surface* | 1000 | 1900 | 2000 |
| 179 to −251 | 2400 | 4500 | 2500 |
| −251 to −851 | 3000 | 5400 | 3000 |
| Below −851 | 3400 | 6000 | 3200 |
| Sediments ($h = 150$ m)† | 600 | 1500 | 2000 |

*There is a velocity gradient in the top layer from the surface down to $z = 179$ m.

†This last line corresponds to the homogeneous sediment filling of the basin.

Figure 1. Map of central Pyrenees, with instrumental and historical seismicity, and the location of the pilot site at Bagnères-deBigorre. NPF denotes the North Pyrenean fault, the boundary of the Iberian and Eurasian plates.

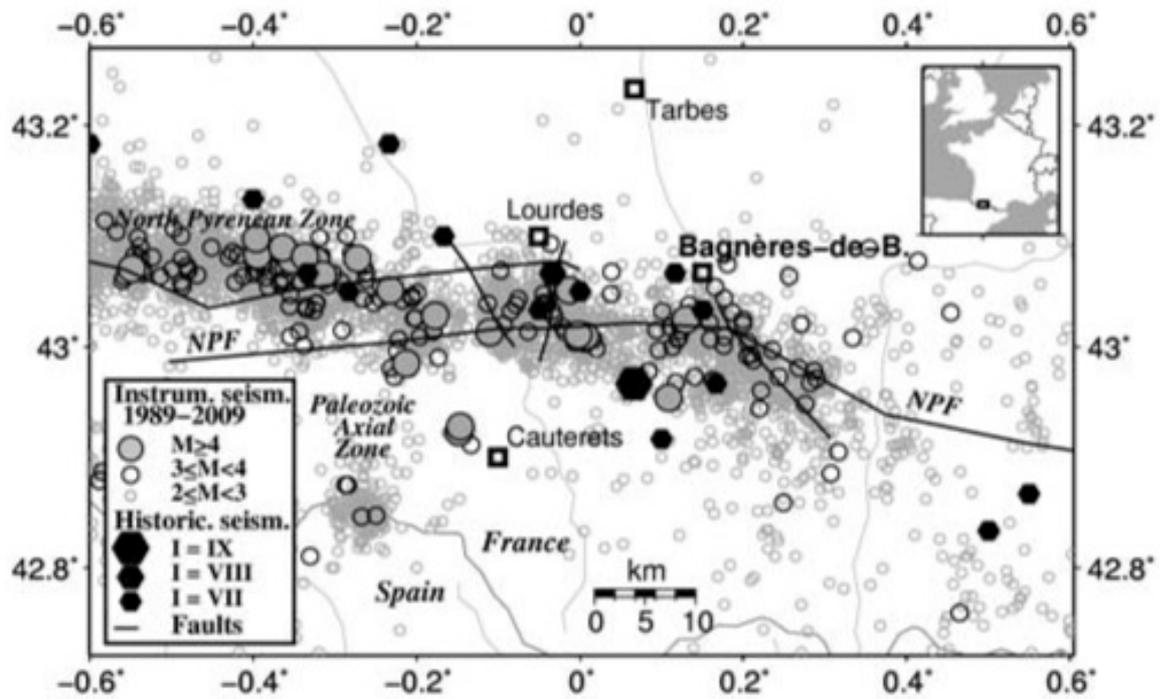

Figure 2. (a) Topographic map of the Adour Valley (from Institut Géographique National), with the location of the accelerometric stations (red dots); (b) geological map of the Adour Valley with the city of Bagnères-de-Bigorre (Azambre et al., 1989). In (a) and (b), the box delineates Fig. 2c; (c) topographic map of the valley, the location of the 10 temporary accelerometric stations, and the permanent accelerometric observatory PYBB. Geological map caption in (b): (1) Paleozoic Ordovician (schist and sandstone), (2) Upper Trias (marls and limestone), (3) Lias (breccias), (4) Dogger and Malm (dolomite, breccias, limestone), (5) Lower Albian (limestone, calcarenites), (6) Mesozoic breccias with granite and gneiss elements, (7) Albian and Cenomanian black flysch, (8) Upper Cretaceous schistous marls, (9) Miocene and Pliocene clay with pebbles, (10) Quaternary fluviatile sediments, (11) Migmatites, (12) Ophites, cmain erosion fans.

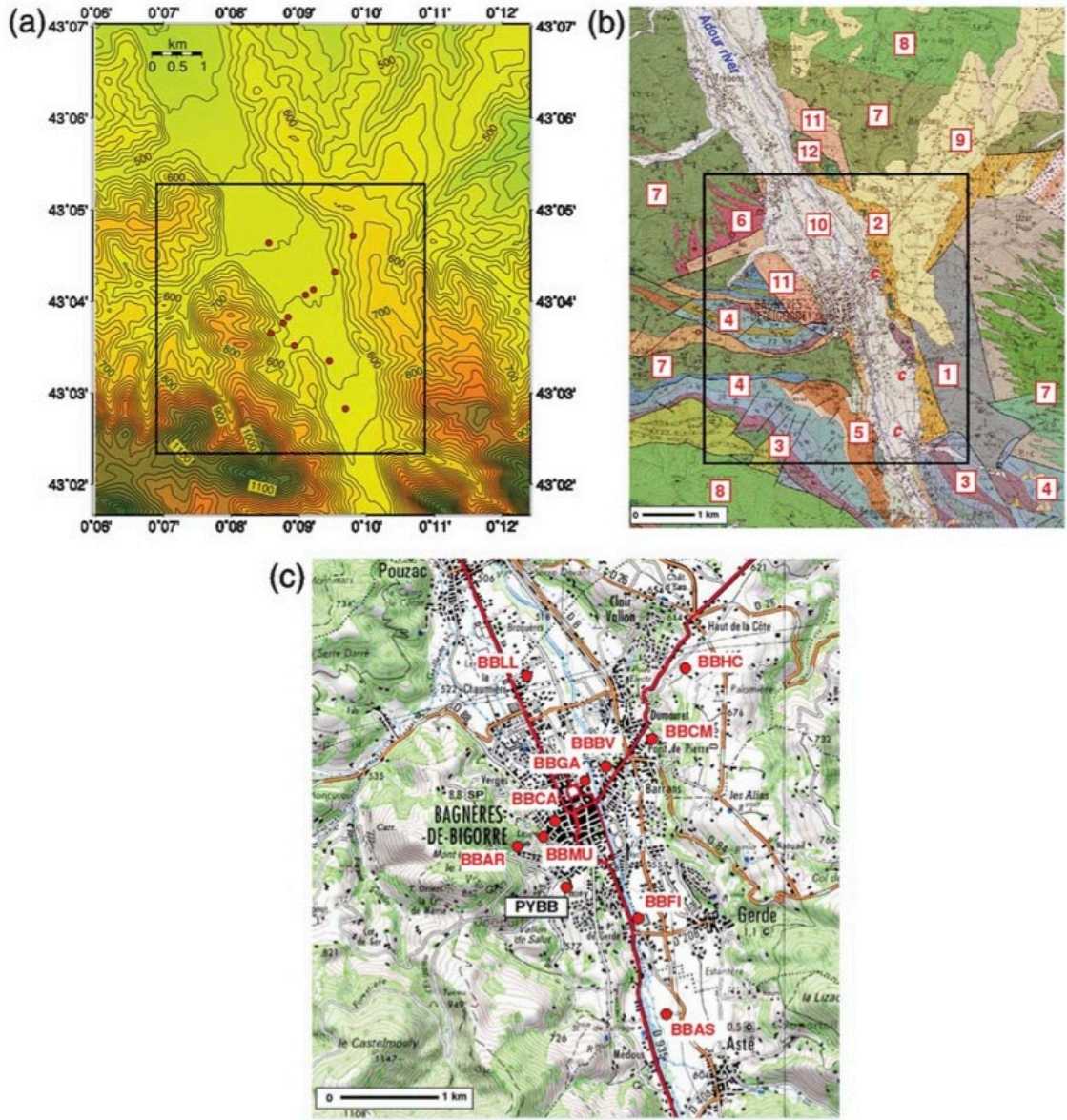

Figure 3. Example of accelerometric records (in black) for the 15 November 2007 event of magnitude 4.3 located 13 km to the west-southwest of Bagnères (east component). Station codes refer to Figure 2c. Ground velocity is plotted in gray to make the low frequencies more apparent. Note the strong variations in amplitude, frequency content, and coda length.

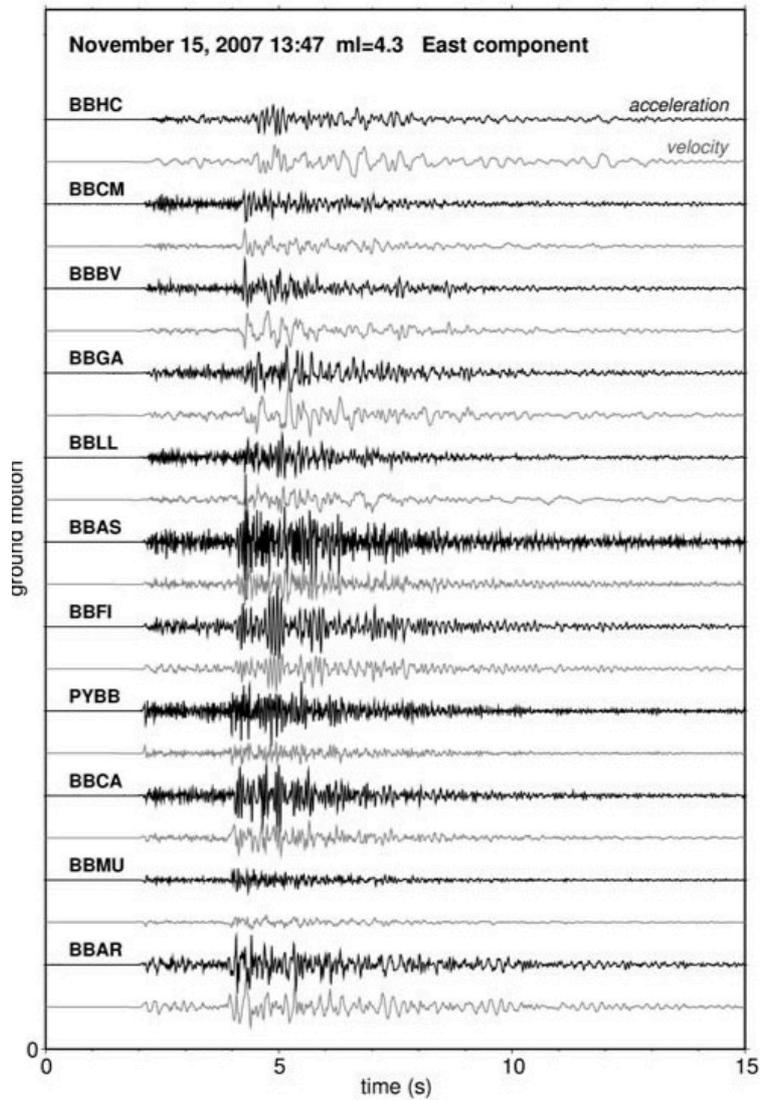

Figure 4. Shear-wave velocity profiles obtained near the sites of the temporary stations from MASW experiments. (a) Inversion of Rayleigh wave dispersion curves, the 1000 best models; (b) simultaneous inversion of dispersion curves and ellipticity (except at BBMU and BBAS, obtained from dispersion curves alone, and BBCM, where no reliable model could be obtained). The thick lines show the mean models; dashed lines delineate the limits of all acceptable models. The velocity profiles can be obtained at greater depth when ellipticity is considered.

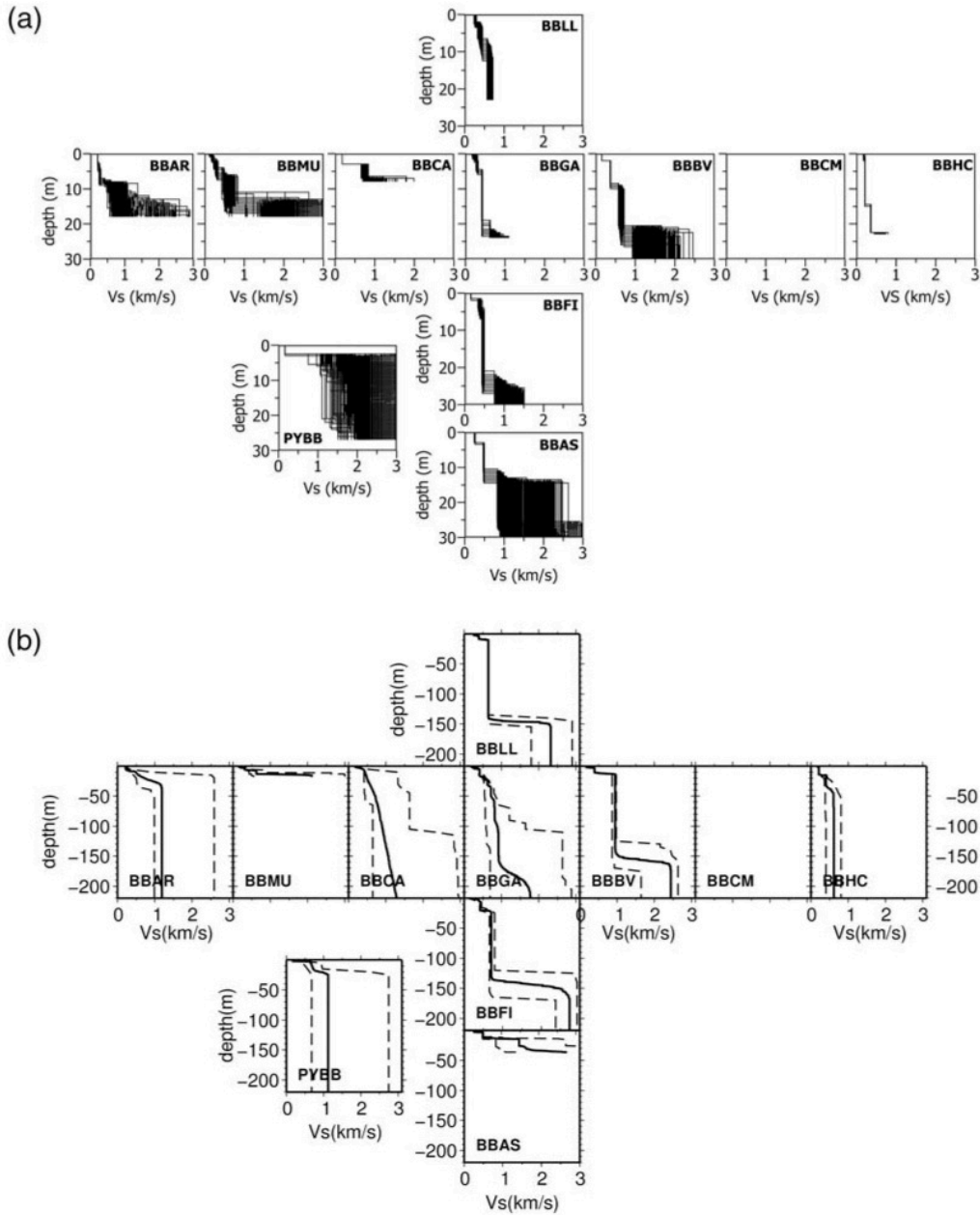

Figure 5. Example of determination of the 1D S-velocity profile from MASW experiment, at site near BBFI. (a) Set of possible structures determined from the phase velocity dispersion curves alone; the fit of the observations (in black) with the model predictions (in gray) is shown in (b) for the dispersion curves (fundamental and first higher mode), and in (c) for the ellipticity of the fundamental mode; (d) structure obtained from simultaneous inversion of phase velocities and ellipticity; comparison with observations is shown in (e) and (f). The introduction of the ellipticity as an additional constraint allows us to retrieve the structure at greater depth.

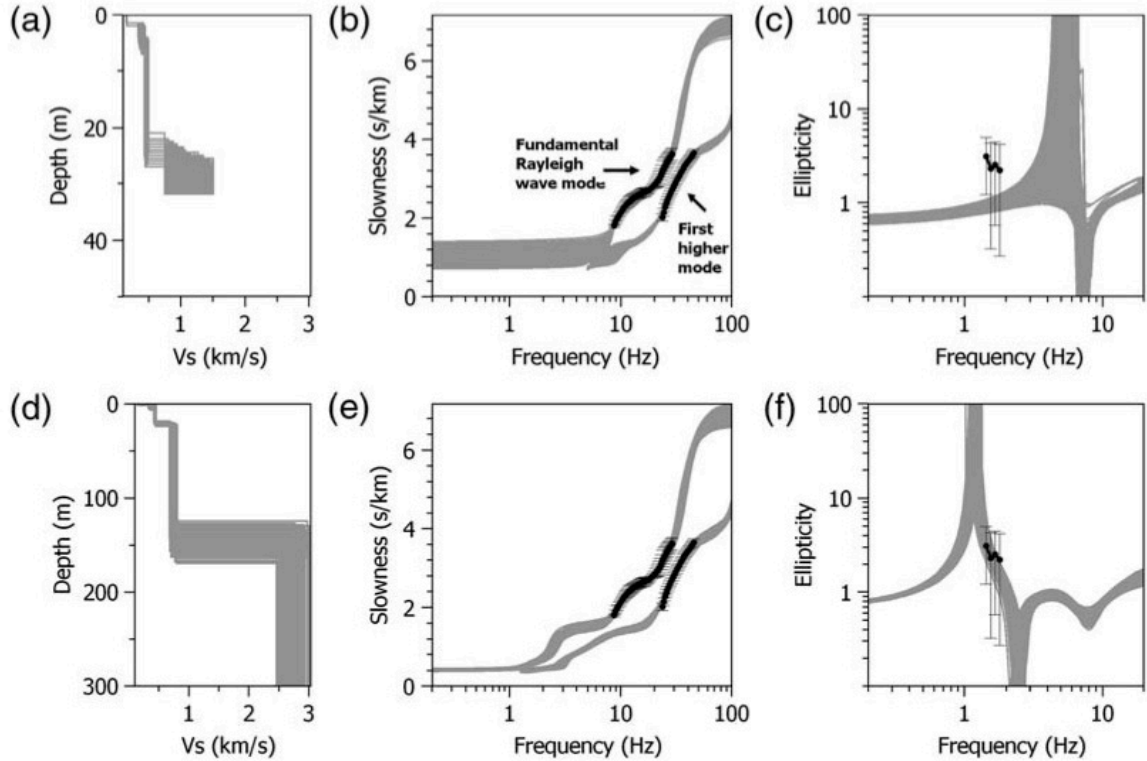

Figure 6. The horizontal-to-vertical spectral ratios on ambient noise. (a) Examples of results obtained at BBCA and BBCM with the accelerometer of the temporary stations (gray), and with the velocimeter of the CityShark instrument (black). For clarity, the confidence level (one standard deviation) is reported for the accelerometer only, it is similar for the velocimeter. (b) Comparison between the power spectral density of the noise at BBCA and the self-noise of the accelerometer. The gray domain is bounded by the low-noise/high-noise curves of Peterson (1993). (c) Influence of the buildings at a few sites. The horizontal-to-vertical ratio is shown in black (with 1σ-confidence level) inside the building and in gray outside the building. (d) Horizontal-to-vertical ratio values with 1σ-confidence level obtained from CityShark at the different stations. Stations denoted by an asterisk (*) are located on rock.

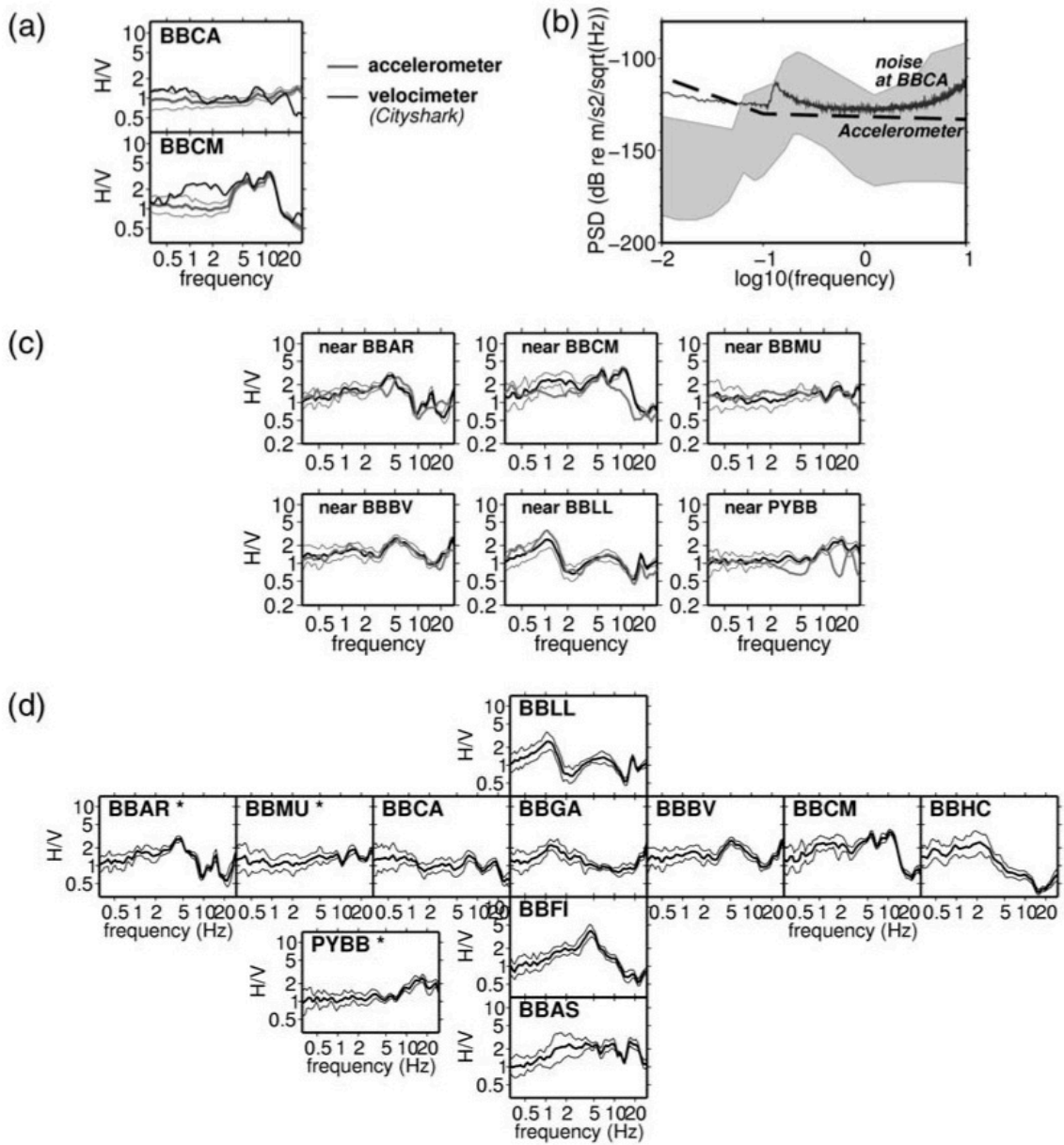

Figure 7. (a) Map of selected sites for H/V measurements on noise with CityShark in the Bagnères Valley; (b) H/V spectral ratios (numbers refer to the sampled locations).

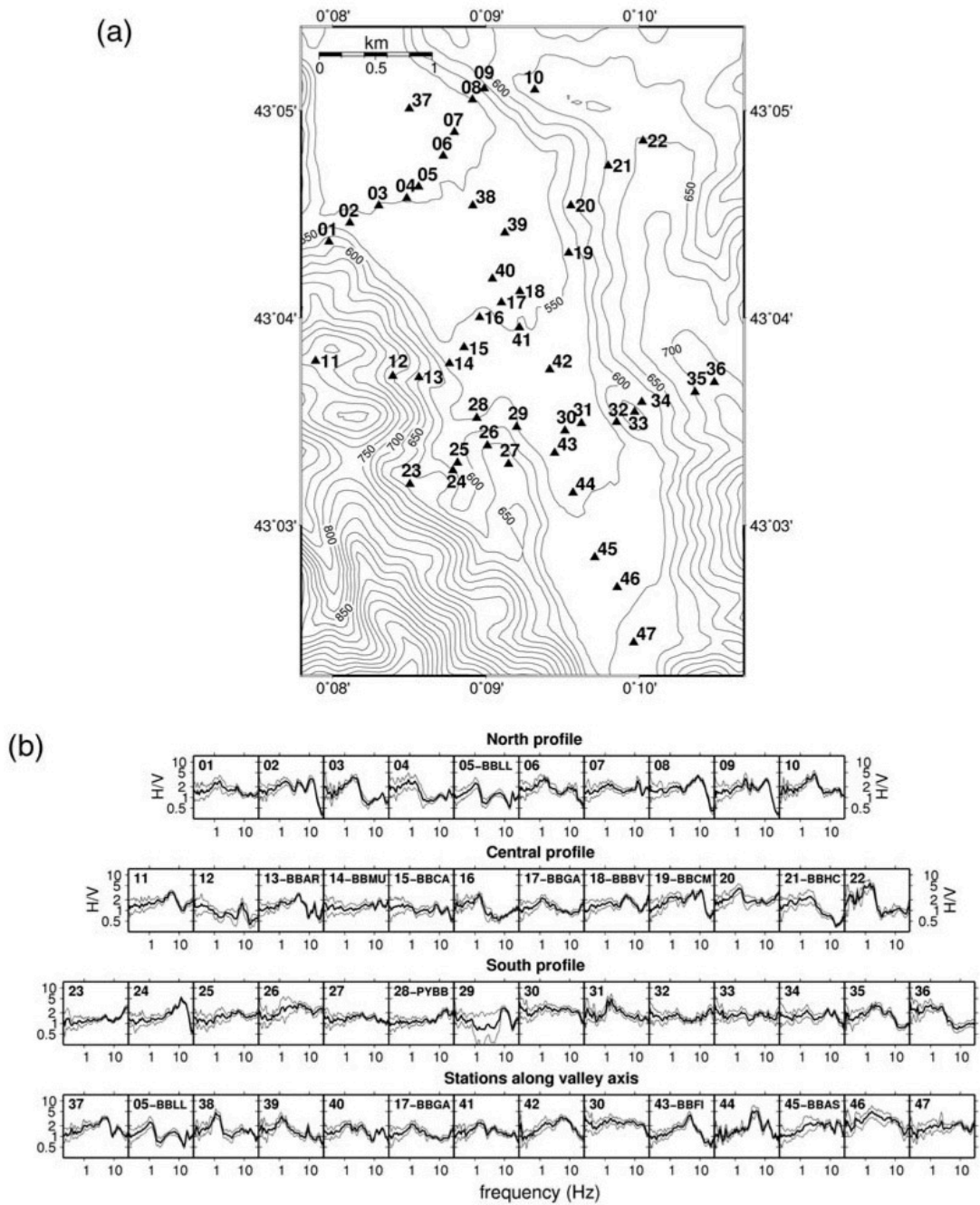

Figure 8. Illustration of equipartition energy in the coda (diffusion regime). (a) Record of the 3 October 2007, $M_L$=2.9 event, located 38 km to the west of Bagnères (east component at BBAR); (b) plot of the seismogram envelope and the mean noise level measured on the first 20 s of the record (dotted line); (c) ratio between the kinetic energy of the vertical and horizontal components, V2/H2, for the record filtered at 15 Hz. Note the stabilization of the ratio between 40 s (i.e., only 4 s after the S-wave arrival) and 100 s. After 100 s, the large fluctuations correspond to the decay of the signal below the noise level.

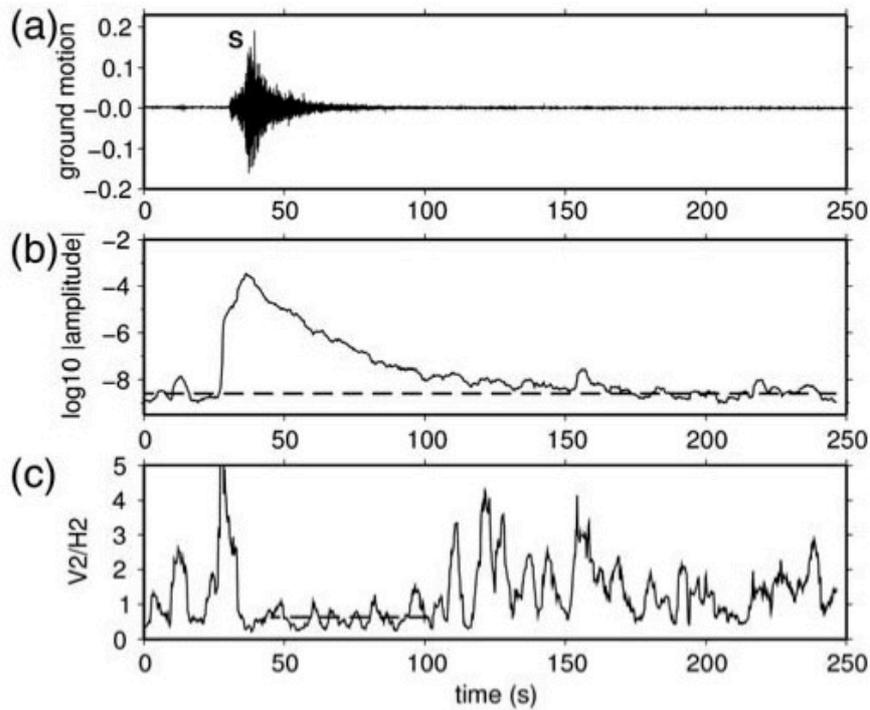

Figure 9. (a) Map of events used for the determination of the H/V ratios on S-coda-waves and the H/Href ratios on S wave and S coda; (b) H/V ratio on S coda (in black), with 1σ-confidence level. The H/V ratios on noise are reported in gray for comparison. Station codes with an asterisk (*) correspond to locations on rock.

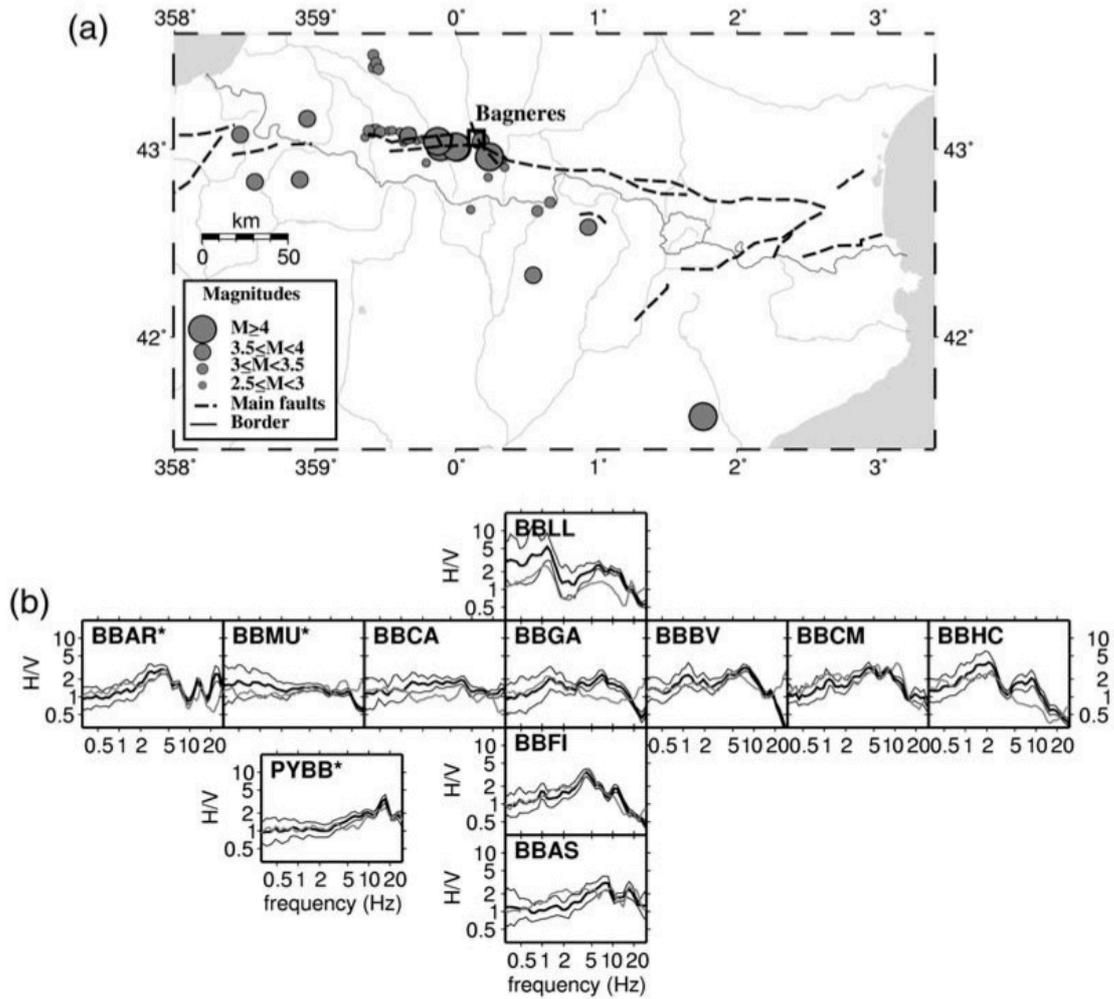

Figure 10. Spectral ratio of horizontal motion (black) and vertical motion (gray) with respect to a reference station (PYBB) (a) for S waves, (b) for the coda of S wave. Events used are shown in Figure 9a. 1σ-confidence level is given for H/Href only, it is similar for V/Vref. In (c) a comparison of the different methods is shown. Station codes with an asterisk (*) correspond to locations on rock.

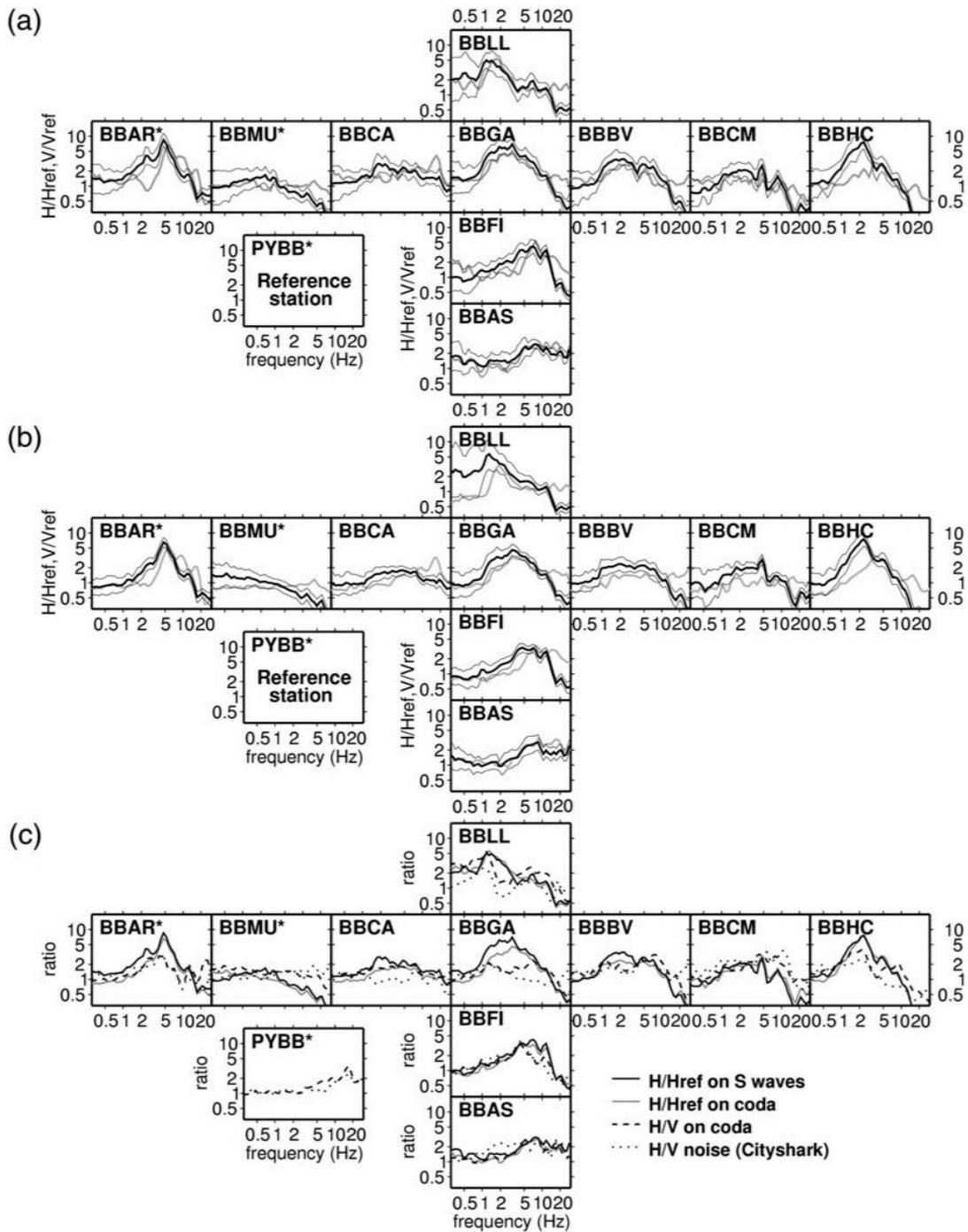

Figure 11. (a) Modeling of the H/V spectral ratio at the temporary stations from noise generated by summing synthetic seismograms. Predicted ratios are in black; experimental spectra of H/V on noise measured outside the buildings are in gray. Values are given with one standard deviation confidence level (thin lines). (b) The uppermost 40 m of the S-velocity models used (defined down to 150 m); only stations with well-defined structure are considered.

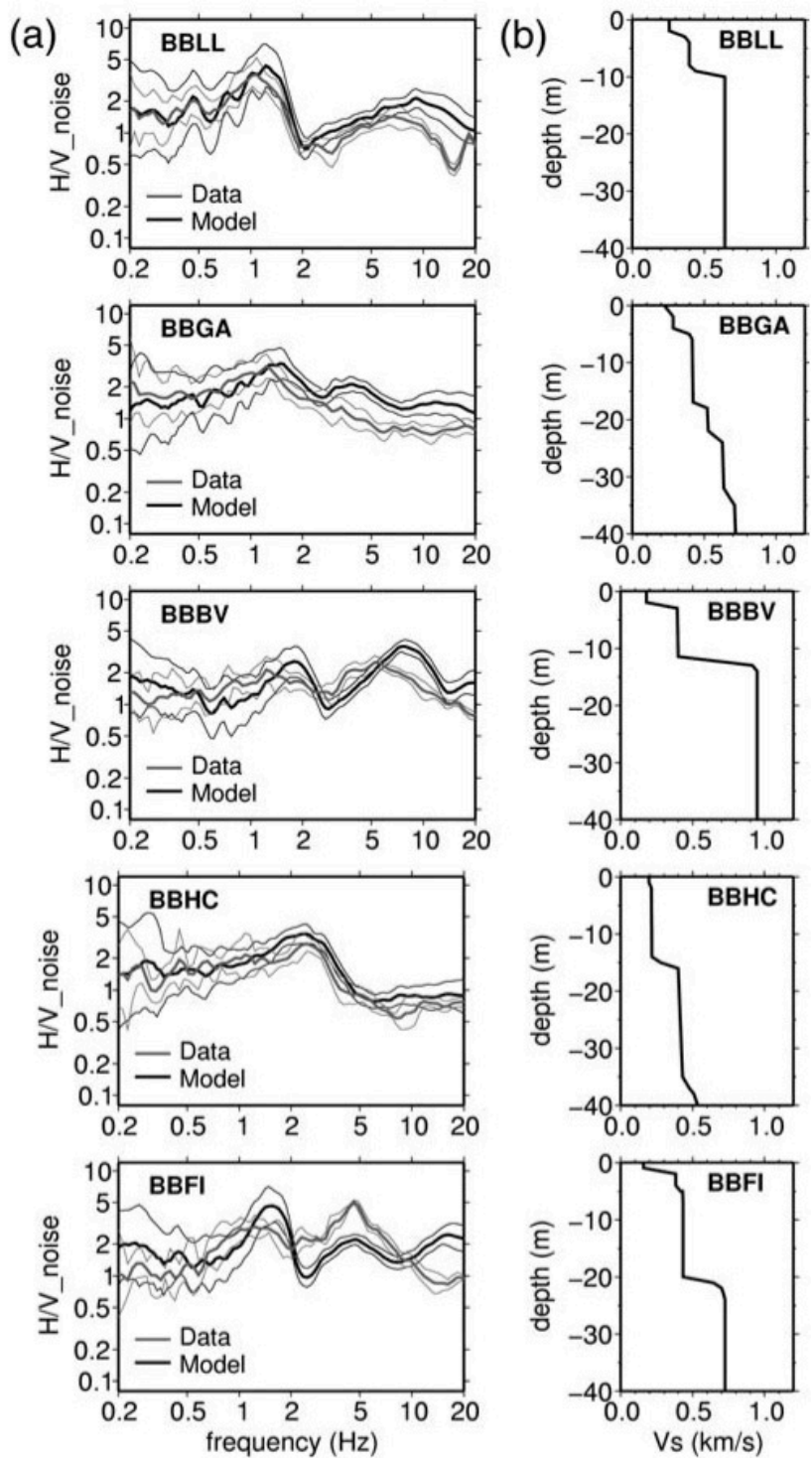

Figure 12. Analysis of energy partitioning in the coda (from the H/V ratio) at the temporary stations. Observed ratios, with 1σ confidence level are in gray. Results of simulations based on equipartition theory with both surface waves and body waves are in black. Structure not available for modeling at BBCM.

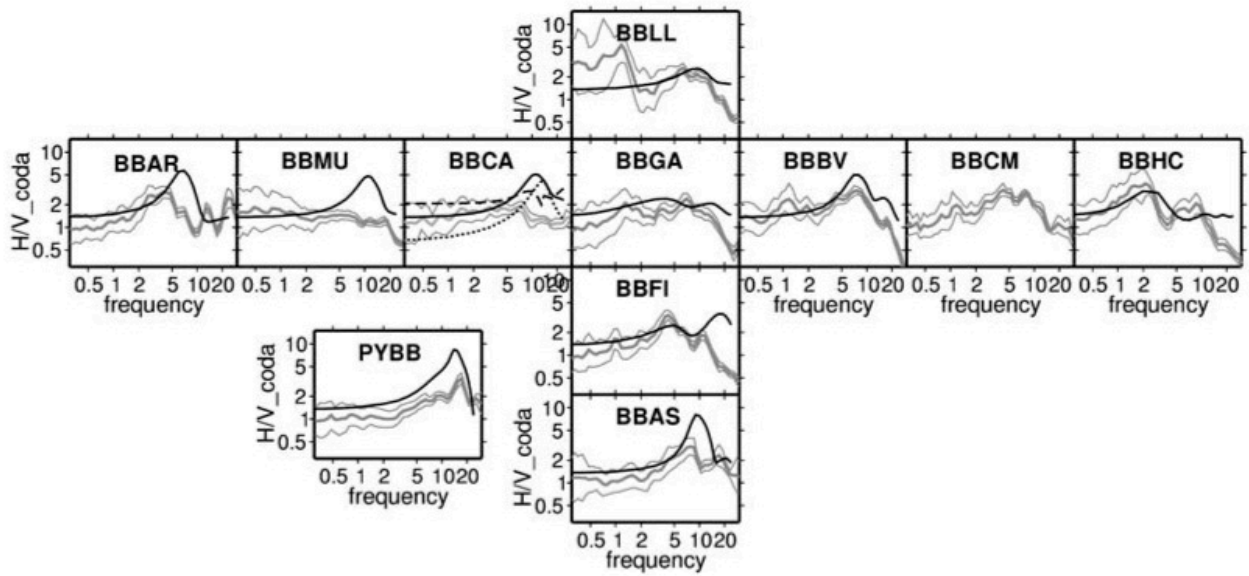

Figure 13. Grid for the spectral element method computations. (a) 3D grid used for the vertically incident plane waves with the location of the basin (in blue); (b) domain of investigation, the white box corresponds to (a); points at the surface indicate where the signal is computed, on a regular grid with spacing 500 m (yellow triangles), and at the sites of the accelerometric stations (red triangles). The domain of investigation is superimposed on the topography in (b). The sedimentary filling is located inside the red contour. The whole map corresponds to the domain considered for modeling the realistic source, whose focal mechanism is plotted at its epicentral location.

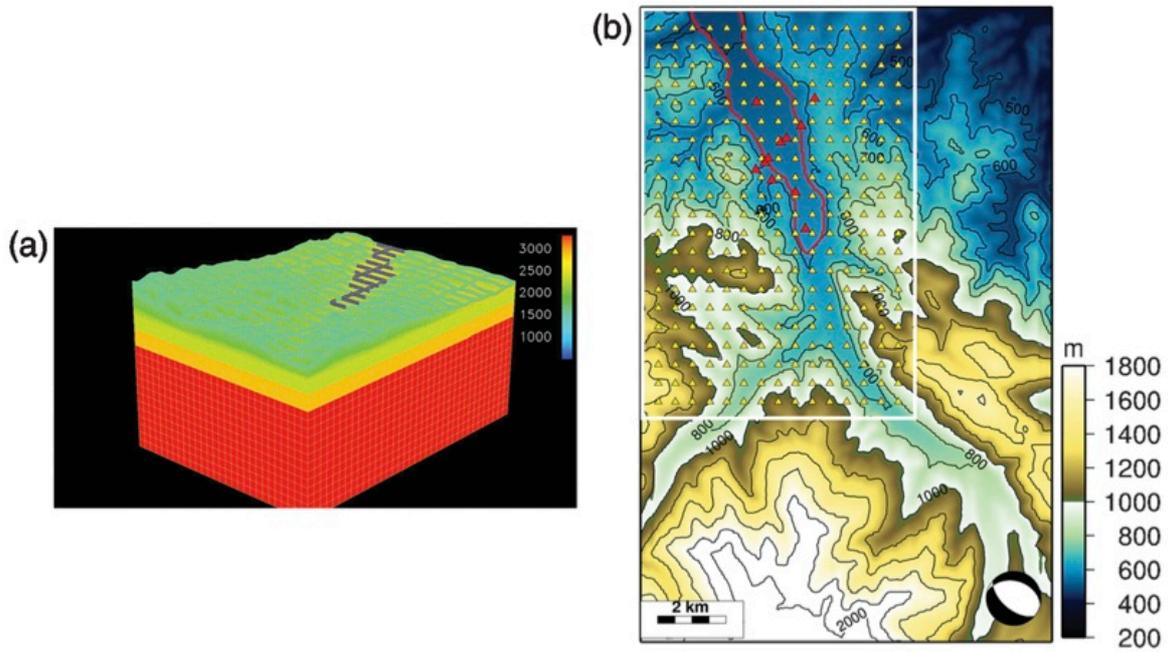

Figure 14. Peak ground acceleration for a plane S-wave incident beneath the structure, with topography alone (a,c) and with the basin filled with sediments (b,d). The location of the basin is shown by a green line. The plane wave has either (a,b) a north–south polarization or (c, d) an east–west polarization. Edge effects are observed along the border of the valley, with an amplitude increase (in red) and an amplitude decrease (in blue) immediately outside. Scale is arbitrary. Numerical artefacts are observed at the border of the plots, over a width of about 2 km.

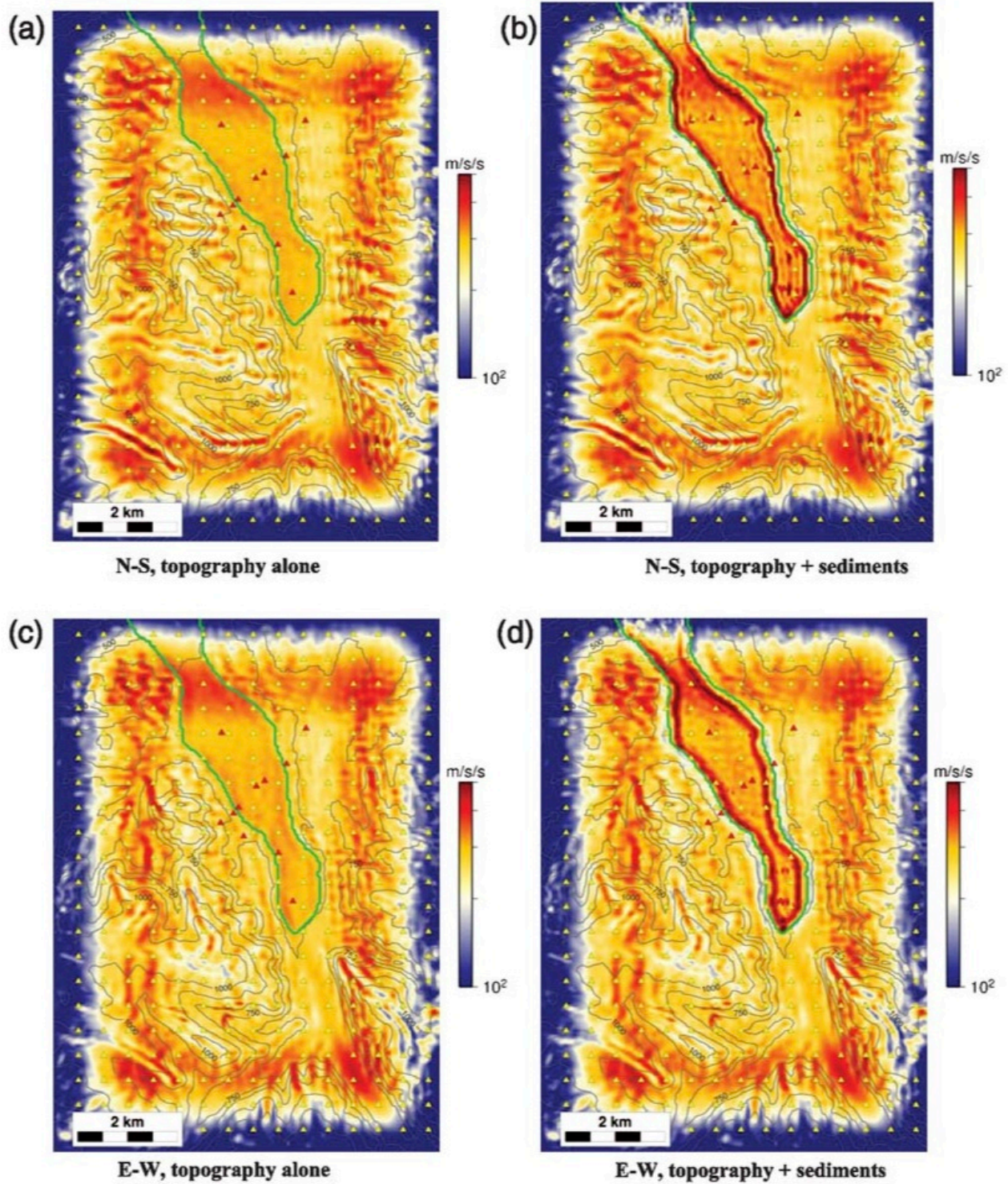

Figure 15. Synthetic ground acceleration for a plane S-wave incident beneath the stations, with north–south polarization. North component at stations of the transverse profile, illustrating the influence of the sediments. (a) Numerical results for topography alone; (b) topography and basin filled with sediments. Note the large amplitudes and long duration inside the basin (BBGA, BBBV), the edge effect appearing as an amplitude decrease immediately outside the basin (BBMU), and as an amplitude increase at the border of the basin (BBCA).

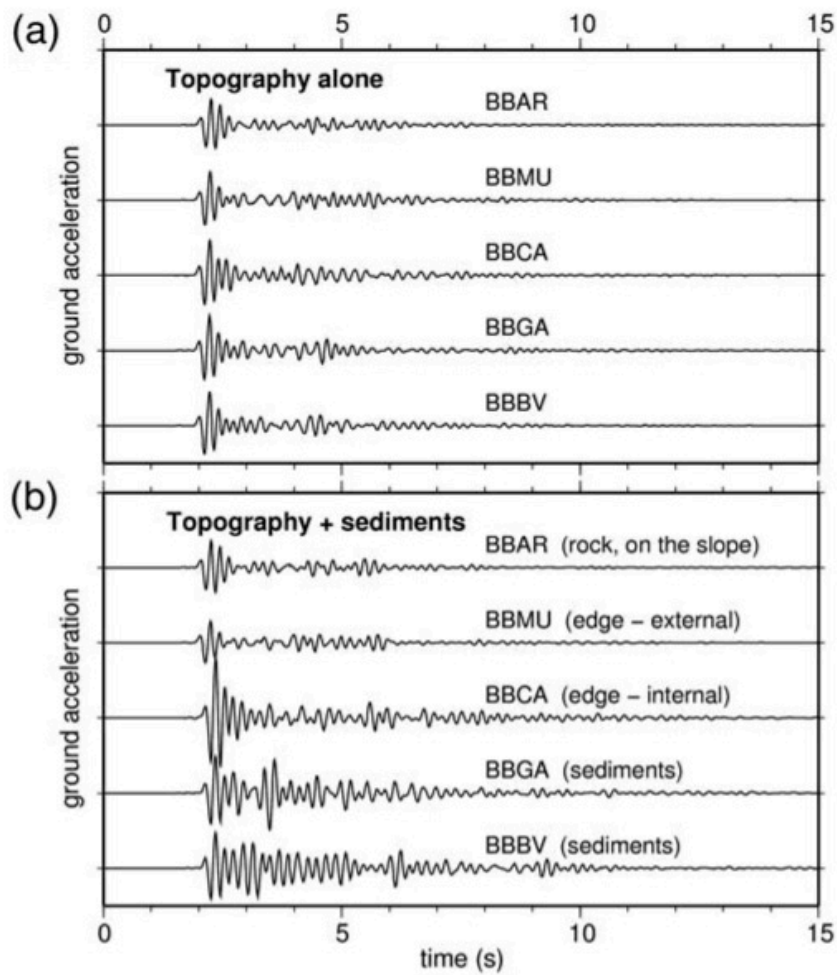

Figure 16. Peak ground acceleration computed for a double-couple point source close to that of the 3 May 2008 event (see location and focal mechanism on Fig. 13b), in a structure with (a) topography alone and (b) topography and sedimentary filling in the basin. Because of the high corner frequency, the absolute PGA values are unrealistic; only the PGA pattern can be analyzed (see text). Note in (a) the predominant influence of radiation pattern on the distribution of PGA, in (b) the general increase of PGA inside the basin and the edge effect with an increase of amplitude inside the basin and a decrease immediately outside. The virtual stations numbered in black correspond to sites considered in Figures 17 and 18.

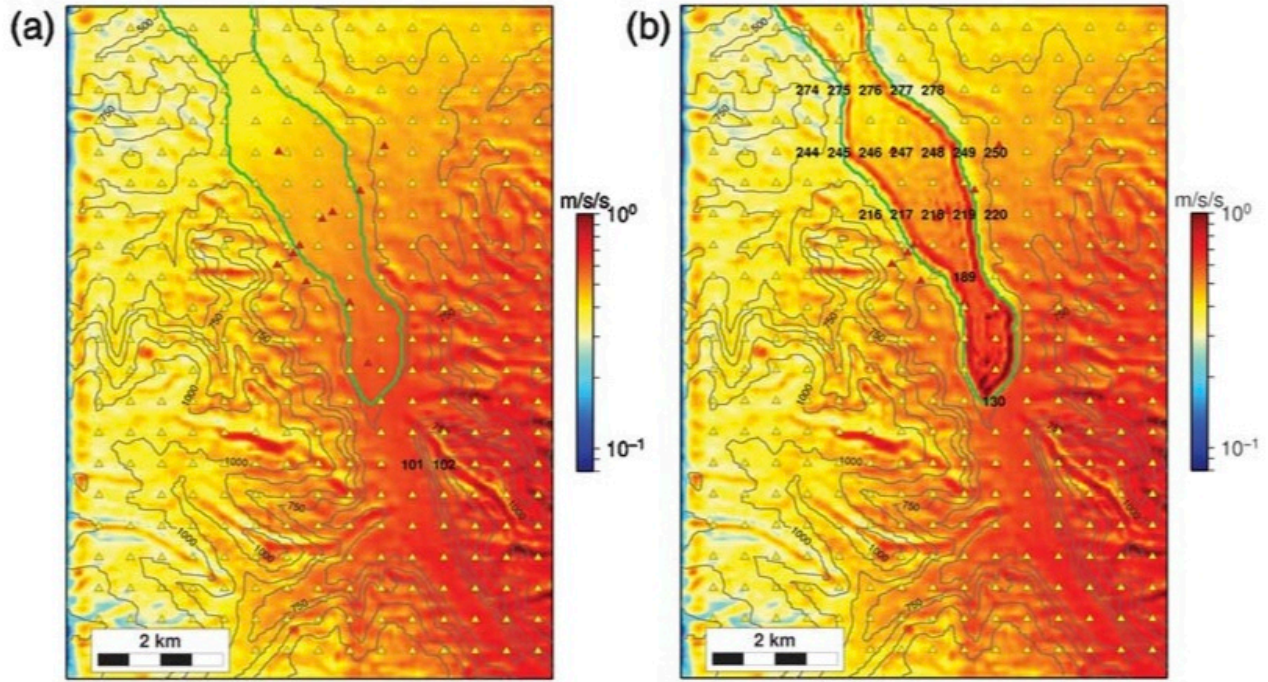

Figure 17. Influence of the sediments on the spectral ratios. Ratios between acceleration spectra obtained from numerical simulations, with topography and sediments, and with topography alone, at selected points (see Fig. 16b), for the double-couple source.

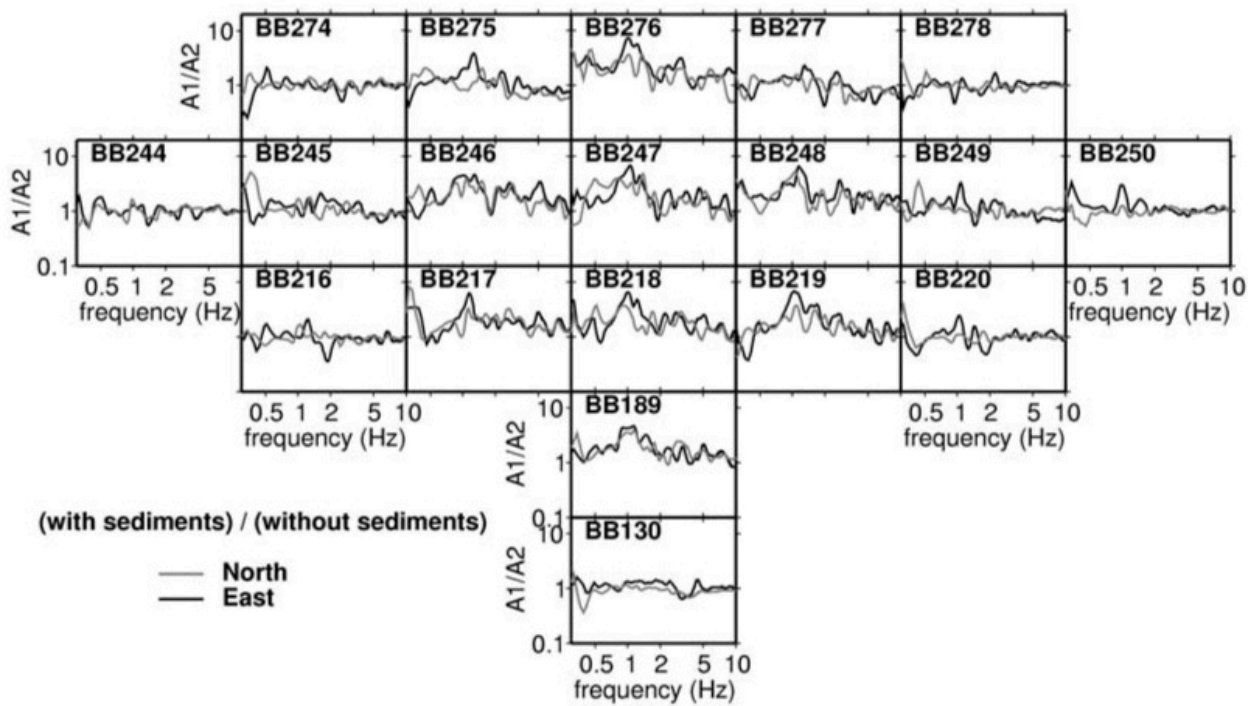

Figure 18. Comparison of the accelerometric records (a) with the synthetics obtained from the 3D-modeling of the valley with the spectral element method. Numerical results for topography alone (c) and for a basin filled with a homogeneous layer of sediments (b). For this last case, results at BBHC and BBCM are not valid (structure unknown). In each panel, the upper, lower, and gray traces correspond to the transverse profile, the longitudinal profile, and the reference station PYBB, respectively. (c) Two synthetic records show the amplification at the top of a crest compared with the hill foot (see station locations and corresponding PGAs on Fig. 16a). The scale is the same for all the traces in (b) and (c) and is normalized to the recorded S-wave amplitude at PYBB.

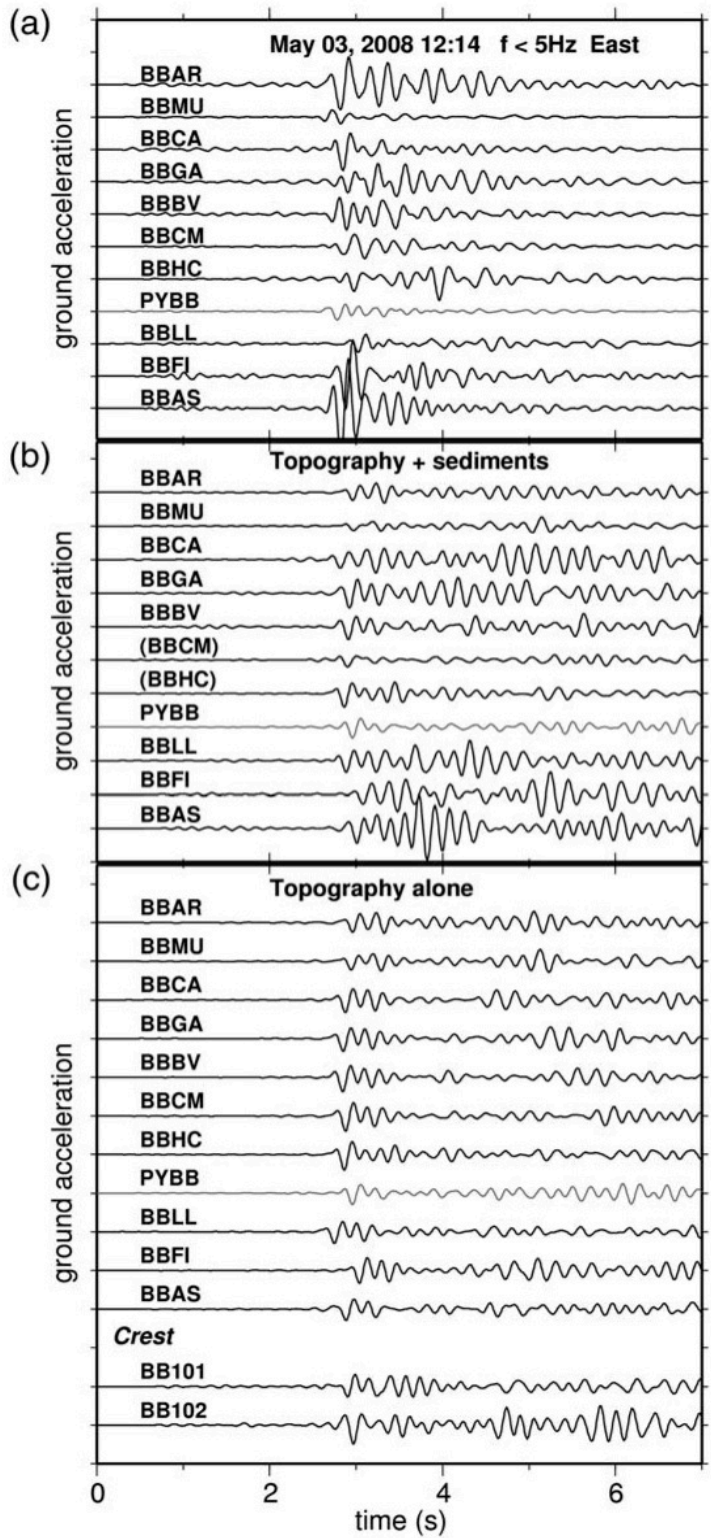